\documentclass[12pt]{article}

\usepackage{latexsym}
\usepackage{epsfig}

\begin{document}
\begin{center}
{\LARGE\bf
Topological susceptibility in full QCD:\\
lattice results versus the prediction from the\\
\vspace{4pt}
QCD partition function with ``granularity''}
\end{center}
\vspace{4pt}

\begin{center}
{\bf Stephan D\"urr}
\vspace{3pt}\\
{\sl Paul Scherrer Institut, Theory Group}\\
{\sl 5232 Villigen PSI, Switzerland}\\
\vspace{1pt}
{\tt stephan.duerr@psi.ch}
\end{center}
\vspace{4pt}

\begin{abstract}
Recent lattice data from CP-PACS, UKQCD, SESAM/TXL and the Pisa group regarding
the quark mass dependence of the topological susceptibility in 2-flavour QCD
are compared to each other and to theoretical expectations. The latter get
specified by referring to the QCD finite-volume partition function with
``granularity'' which accounts for the entropy brought by instantons and
anti-instantons. The chiral condensate in $N_{\!f}\!=\!2$ QCD, if determined by
this method, turns out surprisingly large.
\end{abstract}
\vspace{4pt}

%%%%%%%%%%%%%%%%%%%%%%%%%%%%%%%%%%%%%%%%%%%%%%%%%%%%%%%%%%%%%%%%%%%%%%%%%%%%%%%

\newcommand{\pad}{\partial}
\newcommand{\pas}{\partial\!\!\!/}
\newcommand{\Dsl}{D\!\!\!\!/\,}
\newcommand{\Psl}{P\!\!\!\!/\;\!}
\newcommand{\hqu}{\hbar}
\newcommand{\ovr}{\over}
\newcommand{\til}{\tilde}
\newcommand{\pri}{^\prime}
\renewcommand{\dag}{^\dagger}
\newcommand{\<}{\langle}
\renewcommand{\>}{\rangle}
\newcommand{\gaf}{\gamma_5}
\newcommand{\lap}{\triangle}
\newcommand{\trc}{\rm tr}
\newcommand{\al}{\alpha}
\newcommand{\be}{\beta}
\newcommand{\ga}{\gamma}
\newcommand{\de}{\delta}
\newcommand{\ep}{\epsilon}
\newcommand{\ve}{\varepsilon}
\newcommand{\ze}{\zeta}
\newcommand{\et}{\eta}
\renewcommand{\th}{\theta}
\newcommand{\vt}{\vartheta}
\newcommand{\io}{\iota}
\newcommand{\ka}{\kappa}
\newcommand{\la}{\lambda}
\newcommand{\rh}{\rho}
\newcommand{\vr}{\varrho}
\newcommand{\si}{\sigma}
\newcommand{\ta}{\tau}
\newcommand{\ph}{\phi}
\newcommand{\vp}{\varphi}
\newcommand{\ch}{\chi}
\newcommand{\ps}{\psi}
\newcommand{\om}{\omega}
\newcommand{\psb}{\overline{\psi}}
\newcommand{\etb}{\overline{\eta}}
\newcommand{\psd}{\psi^{\dagger}}
\newcommand{\etd}{\eta^{\dagger}}
\newcommand{\beq}{\begin{equation}}
\newcommand{\eeq}{\end{equation}}
\newcommand{\bdm}{\begin{displaymath}}
\newcommand{\edm}{\end{displaymath}}
\newcommand{\bea}{\begin{eqnarray}}
\newcommand{\eea}{\end{eqnarray}}
\newcommand{\mr}{\mathrm}
\newcommand{\mb}{\mathbf}
\newcommand{\Nf}{N_{\!f}}
\newcommand{\Nc}{N_{\!c}}

%%%%%%%%%%%%%%%%%%%%%%%%%%%%%%%%%%%%%%%%%%%%%%%%%%%%%%%%%%%%%%%%%%%%%%%%%%%%%%%

\hyphenation{topo-lo-gi-cal simu-la-tion gra-nu-lar theo-re-ti-cal
granu-la-ri-ty mini-mum}

%%%%%%%%%%%%%%%%%%%%%%%%%%%%%%%%%%%%%%%%%%%%%%%%%%%%%%%%%%%%%%%%%%%%%%%%%%%%%%%

\section{Introduction}

Quenched QCD is, by many respects, a reasonable approximation to real
(``full'') QCD; in particular for the light meson spectrum this approximation
was found to work so well that a major effort had to be made to drive
statistical errors small enough so that significant differences between the
computed quenched and the experimentally observed (unquenched) spectrum could
be seen \cite{Kanaya:1999sd}.
On the other hand, there are observables which prove highly sensitive on
quenching effects.
An important example of this latter category is the topological susceptibility
which, in the continuum, is defined%
\footnote{Upon including a factor $e^{\mr{i}qx}$ into the definition
(\ref{chidefone}), $\ch$ would pick up an additional dependence on $q^2$. In
this article we shall only be interested in $\ch$ at zero virtuality.},
through \cite{WittenVeneziano}
\beq
\chi=\int\!d^4z\;\pad_\mu\pad_\nu\<T\{k^\mu(x)k^\nu(y)\}\>
\;,
\label{chidefone}
\eeq
where $z\!=\!x\!-\!y$ and
\beq
k^\mu(x)={g^2\ovr16\pi^2}\ep^{\mu\nu\si\rh}A^\mr{a}_\nu(x)
\Big(
\pad_\rh A^\mr{a}_\si(x)-{g\ovr3}f^\mr{abc}A^\mr{b}_\rh(x)A^\mr{c}_\si(x)
\Big)
\label{cherndef}
\eeq
denotes the Chern current which is related to the topological charge density
via
\beq
q(x)=\pad_\mu k^\mu(x)=
{g^2\ovr32\pi^2}G^\mr{a}_{\mu\nu}(x)\til G^{\mr{a}\,\mu\nu}(x)
\qquad\mr{with}\quad \til G^{\mu\nu}\!=\!{1\ovr2}\ep^{\mu\nu\si\rh}G_{\si\rh}
\;.
\label{topchargedef}
\eeq
The topological susceptibility $\ch$ plays a prominent role in QCD, because its
nonvanishing quenched counterpart $\chi_\infty$ explains through the
Witten-Veneziano relation \cite{WittenVeneziano}
\beq
M_{\et'}^2-(2M_K^2-M_\et^2)={2\Nf\ovr F_\pi^2}\ch_\infty
\label{WittenVenezianoFormula}
\eeq
to leading order in $1/\Nc$ why the $\et'$ is heavier than the octet of
pseudoscalar mesons and hence not a pseudo-Goldstone.
The topological susceptibility $\ch$ in the full theory is an interesting
object to study unquenching effects, because it may deviate from the quenched
prediction $\chi_\infty$ by up to 100\%, if the dynamical quarks tend to be
light.
In this respect one should note that the dependence of $\ch\!=\!\ch(\Nf,m)$ on
the number of dynamical quarks and their masses is not explicit in the
definition (\ref{chidefone}); it enters {\em implicitly\/} via the vacuum
expectation value $\<...\>$ which includes the weight of the fermion functional
determinant.
This means that the topological susceptibility is useful to study the QCD
vacuum structure and how the latter is affected by the presence of light
dynamical quarks.
It is therefore little surprise that the first reliable lattice data for
the topological susceptibility in full QCD were eagerly awaited for.

Below we shall analyze the data by CP-PACS \cite{CPPACS}, UKQCD \cite{UKQCD},
SESAM/TXL \cite{SESAM} and the Pisa group \cite{PISA}.
We will see that they are roughly consistent with each other, but follow
theoretical expectations to a limited extent only.
Potential sources of error are analyzed, and it is shown that specifically for
the topological susceptibility finite-volume effects may be quantitatively
assessed by deriving it from the QCD finite-volume partition function for which
an elaborate version is given.
On the practical side, a non-standard determination of the chiral condensate in
$\Nf\!=\!2$ QCD is presented, and the result is found surprisingly large.

%%%%%%%%%%%%%%%%%%%%%%%%%%%%%%%%%%%%%%%%%%%%%%%%%%%%%%%%%%%%%%%%%%%%%%%%%%%%%%%

\section{Theoretical expectations}

To begin, we shall remind ourselves of the theoretical expectations regarding
the dependence of the topological susceptibility $\ch$ on the sea-quark masses
$m_1,\ldots,m_{\Nf}$.
For simplicity the latter shall be taken degenerate, i.e.\ we are interested
in the function $\ch\!=\!\ch(\Nf,m)$ with $\Nf\!\geq\!2$.
In the present section the four-volume $V$ of the box is assumed ``infinite'';
finite-volume effects will be analyzed below.

Detailed predictions on $\ch\!=\!\ch(\Nf,m)$ may be obtained for the two
regimes where the quark masses tend to be either very light or very heavy.
In the former case the flavour singlet axial Ward-Takahashi (WT) identity
dictates that $\ch$ raises {\em linearly\/}%
\footnote{Here the assumption $V\!\to\!\infty$ proves essential; for the
behaviour at finite $V$ see below.}
\cite{Crewther:1977ce, LeutwylerSmilga, Aoki}, viz.\
\beq
\ch={\Sigma\ovr\Nf}\,m+o(m)
\;,
\label{susclarge}
\eeq
where $\Sigma\!=\!-\lim_{m\to0}\lim_{V\to\infty}\<\psb\ps\>$ is the chiral
(one-flavour) condensate in the chiral limit%
\footnote{Both $\Sigma$ and $m$ are scheme- and scale-dependent; the product,
however, is an RG-invariant quantity. In this article, individual quotes for
$\Sigma,m$ mean that standard conventions ($\overline{\mr{MS}},
\mu\!=\!2\,\mr{GeV}$) have been adopted.}
and the $o(m)$ term contains higher order corrections%
\footnote{The reason why I write $o(m)$ rather than $O(m^2)$ is that the higher
order terms may contain chiral logarithms; a chiral expansion is (in general)
non-analytic.}
which may be worked out explicitly in chiral perturbation theory \cite{SXPTrev}.
On the other hand, for heavy dynamical quarks the topological susceptibility
curve will be {\em almost flat\/}, because it must gradually approach the
quenched value $\ch_\infty$ which is finite.
A numerical estimate of the latter quantity by means of the Witten-Veneziano
relation (\ref{WittenVenezianoFormula}) yields
$\ch_\infty\!\simeq\!(180\,\mr{MeV})^4\!\simeq\!1.05\;10^9\,\mr{MeV}^4$.
In addition, it is easy to come up with plausible arguments that $\ch$
{\em grows monotonically\/} as the quark mass increases and that the
{\em second derivative is negative\,}${}^2$.

\begin{figure}%1
\epsfig{file=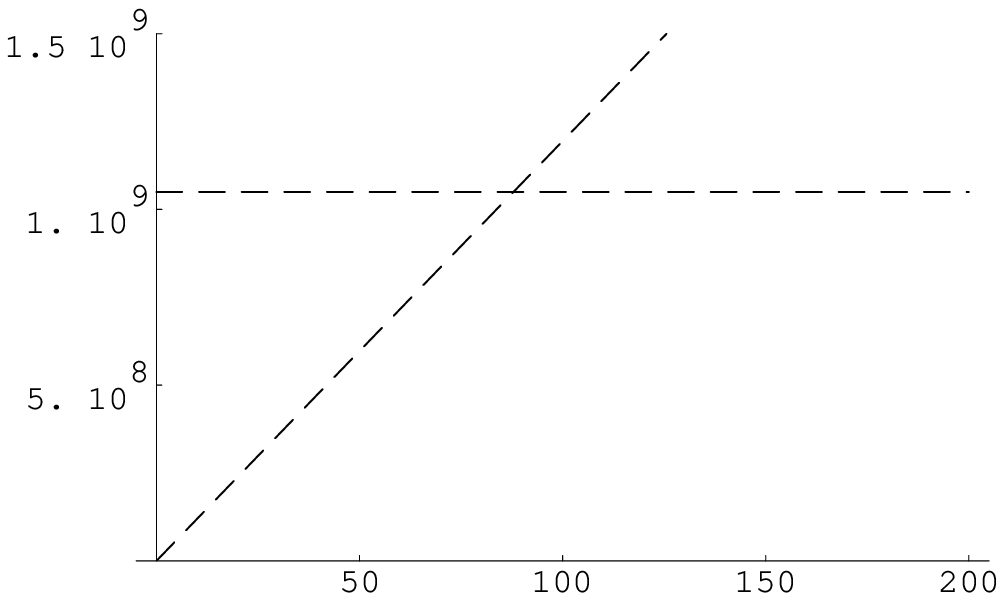,
height=5cm,width=7cm,angle=0}
\hfill
\epsfig{file=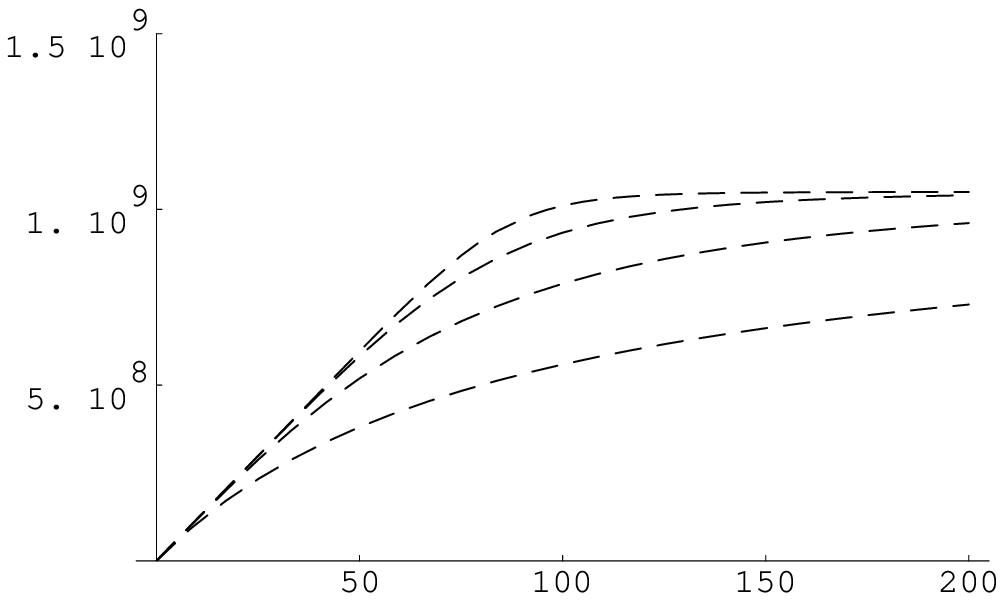,
height=5cm,width=7cm,angle=0}
\vspace{-3mm}
\caption{\sl\small
Basic illustrations regarding the topological susceptibility [in $\mr{MeV}^4$]
versus the quark mass [in $\mr{MeV}$] for 2-flavour QCD with
$\Sigma\!=\!(288\,\mr{MeV})^3$ (cf.\ footnote 5). LHS: The leading order chiral
prediction in the large Leutwyler-Smilga regime without instanton effects
{\rm (\ref{susclarge})} (raising slope) and the quenched prediction from the
Witten-Veneziano relation {\rm (\ref{WittenVenezianoFormula})} (flat line).
RHS: The curves with $n\!=\!1, 2, 4, 8$ (from the lowest to the highest) out of
the phenomenological family {\rm (\ref{chipheno})}, all of which share the
required asymptotic behaviour.}
\end{figure}

Hence, without doing a detailed analysis, a clear picture emerges for the
{\em qualitative\/} behaviour of the topological susceptibility as a function
of the quark mass over the whole range of the latter.
Even the transition regime may be estimated by just equating the leading-order
chiral expression (\ref{susclarge}) to the asymptotic value $\ch_\infty$, as
is illustrated on the l.h.s.\ of Fig.\ 1:
Upon using%
\footnote{The numerical value stems from the Gell-Mann--Oakes--Renner (GOR)
PCAC-relation $M_\pi^2 F_\pi^2\!\simeq\!2\Sigma m$, together with
$M_\pi\!\simeq\!139\,\mr{MeV}, F_\pi\!\simeq\!93\,\mr{MeV}$ and
$m\!\simeq\!3.5\,\mr{MeV}$ (cf.\ footnote 3) \cite{AliKhan:2000mw}.}
$\Sigma\!=\!(288\,\mr{MeV})^3\!=\!2.4\;10^7\,\mr{MeV}^3$ one gets
$m_\mr{trans}\!\simeq\!88\,\mr{MeV}$ for 2-flavour QCD, which is accessible in
current dynamical simulations.

What these considerations cannot yield is the {\em functional form\/} of the
topological susceptibility curve.
Of course it is easy to come up with a whole bunch of phenomenological models
or ans\"atze which interpolate in various ways between the asymptotic regimes;
e.g.\
\beq
\ch(\Nf,m)=((\Sigma\;m/\Nf)^{-n}+\ch_\infty^{-n})^{-1/n}
\label{chipheno}
\eeq
is a one-parameter family of interpolating curves (some of which are displayed
in the r.h.s.\ of Fig.\ 1) with the required asymptotic behaviour.

In summary, one would expect up-to-date simulations of full 2-flavour QCD to
show the turn-over from a linear to an almost flat behaviour at a quark mass
of $O(88\,\mr{MeV})$, and to select the most promising phenomenological overall
fitting curve from the set (\ref{chipheno}).

%%%%%%%%%%%%%%%%%%%%%%%%%%%%%%%%%%%%%%%%%%%%%%%%%%%%%%%%%%%%%%%%%%%%%%%%%%%%%%%

\section{Lattice data}

Before considering the actual data, we should remind ourselves that it is not
the original definition (\ref{chidefone}) of the topological susceptibility
which is typically used on the lattice, but a simplified version which --~at
least in the continuum~-- proves equivalent.

The simplification takes place as the result of a two-step procedure.
The first step is to assume {\em translation invariance\/} of the vacuum,
so (\ref{chidefone}) simplifies to
\beq
\chi=\int\!d^4x\;\<T\{q(x)q(0)\}\>
\;,
\label{chideftwo}
\eeq
where we have also assumed that the $\de$-functions due to the derivatives in
(\ref{chidefone}) hitting the $T$-ordering do not contribute after integration.
The second step assumes that there is, in addition, {\em no correlation\/}
between topological charge densities, so (\ref{chideftwo}) simplifies to
\beq
\chi={\<q^2\>\ovr V}
\;,
\label{chidefthr}
\eeq
where
\beq
q={g^2\ovr32\pi^2}\int\!d^4x\;G^\mr{a}_{\mu\nu}\til G^{\mr{a}\;\mu\nu}
\label{chargedef}
\eeq
is the (global) topological charge which, in the continuum, takes an integer
value.
The important point is that the original definition (\ref{chidefone}) provides
a detailed prescription of how to deal with the $\de$-like singularity in the
product $q(x)q(0)$ which is not explicit any more in the simplified version
(\ref{chidefthr}).
As a matter of consequence \cite{TopChargeRenormalization}, a lattice
implementation of (\ref{chidefone}) is subject to a {\em multiplicative\/}
renormalization only
\beq
\hat\ch=Z^2\,a^4\,\ch+O(a^6)
\label{chirenmul}
\eeq
with the renormalization factor of the topological charge
\beq
\hat q=Z\,q+O(a^2)
\;,
\label{topren}
\eeq
whereas a lattice version of (\ref{chidefthr}) shows an additional
{\em additive\/} renormalization, viz.
\beq
\hat\ch=M(\be,\Nf,m)+Z^2\,a^4\,\ch+O(a^6)
\,.
\label{chirenadd}
\eeq
While in practice both the multiplicative renormalization factor $Z$ and the
additive piece $M$ do not cause any difficulties, since they may be determined
by means of heating techniques \cite{PisaHeating}, an explicit verification of
the three definitions (\ref{chidefone}, \ref{chideftwo}, \ref{chidefthr})
proving equivalent after they have been implemented on the lattice is still
being awaited.
Since, as we have seen, proper treatment of contact terms is essential in
this game (cf.\ \cite{Vicari:1999xx} and references therein), such an
investigation might be interesting in its own right.

\begin{figure}[t]%2
\begin{center}
\vspace{-14mm}
\epsfig{file=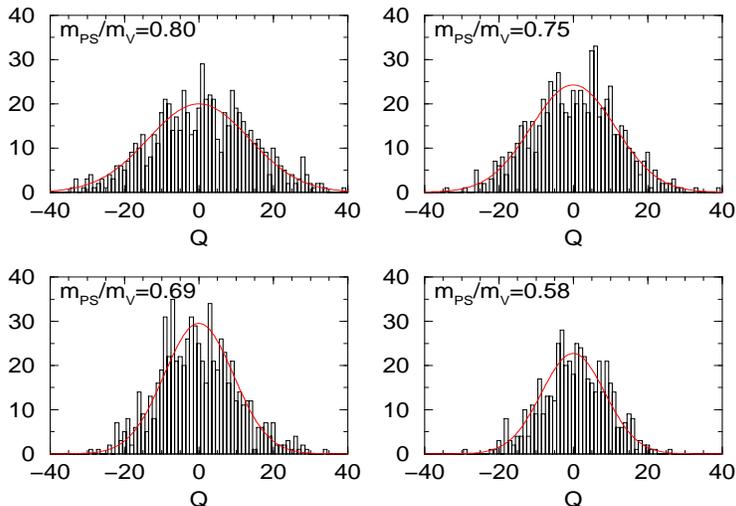,
height=8.8cm,width=12.0cm,angle=0}
\vspace{-16mm}
\end{center}
\caption{\sl\small
CP-PACS data for the distribution of topological charges on a
$16^3\!\times\!32$ lattice after cooling with an RG-improved action.
The configurations were generated using the same (Iwasaki) action at fixed
$\be\!=\!1.95$ together with 2 flavours of mean-field improved clover quarks
($c_{{}_\mr{SW}}\!=\!1.53$) at 4 different $\ka$-values.
Figure by courtesy of the CP-PACS collaboration.}
\end{figure}

After these cautionary remarks we are ready to esteem the latest lattice data
which use the simplified definition (\ref{chidefthr}), thereby dealing --~in
principle~-- with the renormalization pattern (\ref{chirenadd}).
In order to get a feeling of what is actually done, it is useful to consider
the distribution of topological charges as found by the CP-PACS collaboration 
\cite{CPPACS}.
Fig.\ 2 shows the histograms they have gotten on a $16^3\!\times\!32$ lattice
at $\be\!=\!1.95$ (using the Iwasaki action) and at four different $\ka$-values
($\ka\!=\!0.1375, 0.1390, 0.1400, 0.1410$, giving $M_\pi/M_\rh\!=\!0.80, 0.75,
0.69, 0.58$, respectively) for 2 flavours of mean-field improved clover quarks
($c_{{}_\mr{SW}}\!=\!1.53$) after cooling with the Iwasaki action and binning
$q$ to integer values.

\begin{figure}[t]%3
\vspace{-3mm}
\begin{center}
\epsfig{file=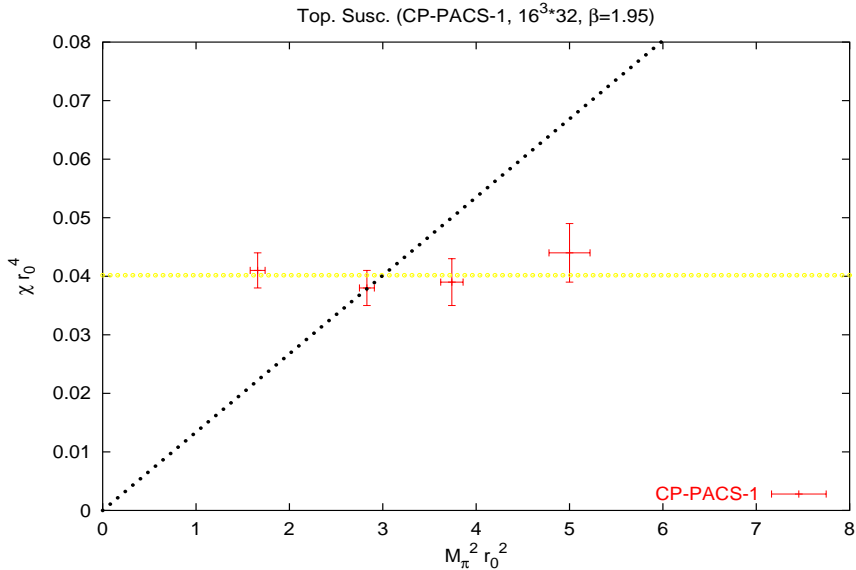,
height=7.7cm,width=11.7cm,angle=0}
\end{center}
\vspace{-7mm}
\caption{\sl\small
CP-PACS data for the topological susceptibility [in units of $r_0^{-4}$] versus
the quark mass [expressed in terms of $M_\pi^2 r_0^2$] in 2-flavour QCD, as
determined from the $q$-distributions shown in Fig.\ 2.
The linearly raising slope indicates the asymptotic behaviour for light
dynamical quarks ($\ch_0 r_0^4\!\simeq\!(F_\pi^2 r_0^2/2\Nf) M_\pi^2 r_0^2\!
\simeq\!0.0133\,M_\pi^2 r_0^2$ for $F_\pi\!\simeq\!93\,\mr{MeV}$ and $r_0\!
\simeq\!0.49\,\mr{fm}$), and the horizontal line the one for (infinitely)
heavy dynamical quarks ($\ch_\infty r_0^4\!\simeq\!(180\,\mr{MeV})^4\,(0.49\,
\mr{fm})^4\!\simeq\!0.04$), as in the l.h.s.\ of Fig.\ 1.
The data are expected to lie in the sector beneath either one of these lines,
i.e.\ $\ch r_0^4\!\leq\!\mr{min}(\ch_0 r_0^4, \ch_\infty r_0^4)$, where
$\ch_\infty$ might be somewhat larger (cf.\ text).
Data taken from \cite{CPPACS,Aoki}.}
\end{figure}

Since the cooling procedure brings both renormalization constants ($Z$ and $M$)
close to their continuum values ($1$ and $0$, respectively), one can basically
read off the expectation values $\<q^2\>$ from the widths of the distributions
shown in Fig.\ 2.
This holds, because
\beq
\<q^2\>\equiv
{\int\!dq\;q^2\;e^{-q^2/2\si^2}
\ovr\int\!dq\;e^{-q^2/2\si^2}}
=\si^2
\label{simple}
\eeq
for a gaussian distribution.
Hence, from a glimpse at Fig.\ 2 one might get the impression that the
topological susceptibility gets {\em reduced\/} as the quark mass decreases,
since the $q$-distribution does indeed get narrower as $\ka$ increases.
What this naive consideration neglects is that the topological susceptibility
$\ch$ is defined as the variance of these gaussian curves divided by the
physical four-volume of the box (cf.\ (\ref{chidefthr},~\ref{simple})), and
the latter changes with $\ka$, because in full QCD the physical lattice spacing
$a$ is a non-trivial function of $\be$ and $\ka$.
To be specific: $a$ decreases if $\ka$ gets increased at fixed $\be$, and for
the volume the reduction is particularly strong, as it involves the fourth
power of $a$.
Taking this effect into account yields the data for $\ch$ versus the dynamical
quark mass $m$ in $\Nf\!=\!2$ QCD shown in Fig.\ 3.
They seem to indicate an {\em almost flat\/} topological susceptibility curve,
and this means that the narrowing of the $q$-distribution in Fig.\ 2 was
exclusively due to the volume $V$ decreasing with increasing $\ka$ (at fixed
$\be$) and not due to a decrease of $\ch$.
In other words: The data in Fig.\ 3 show no sign of the expected transition to
the linear behaviour (\ref{susclarge}) characteristic for light dynamical
quarks, even though the most chiral point is at $M_\pi^2 r_0^2\!=\!1.66$, i.e.\
at a quark mass of $m\!\simeq\!50\mr{MeV}$ (using $F_\pi\!=\!93\,\mr{MeV},
\Sigma\!=\!(288\,\mr{MeV})^3$) which is well below our estimate of the
transition regime; we had $m_\mr{trans}\!\simeq\!88\,\mr{MeV}$ for $\Nf\!=\!2$.
The troublesome point is the most chiral one; it seems to violate the
constraint $\ch\!\leq\!\ch_0$ with $\ch_0$ defined as the r.h.s.\
of~(\ref{susclarge}).
Potential reasons for the supposed excess of the measured $\hat\ch$ will be
discussed below.

\begin{figure}[p]%4
\vspace{-4mm}
\begin{center}
\epsfig{file=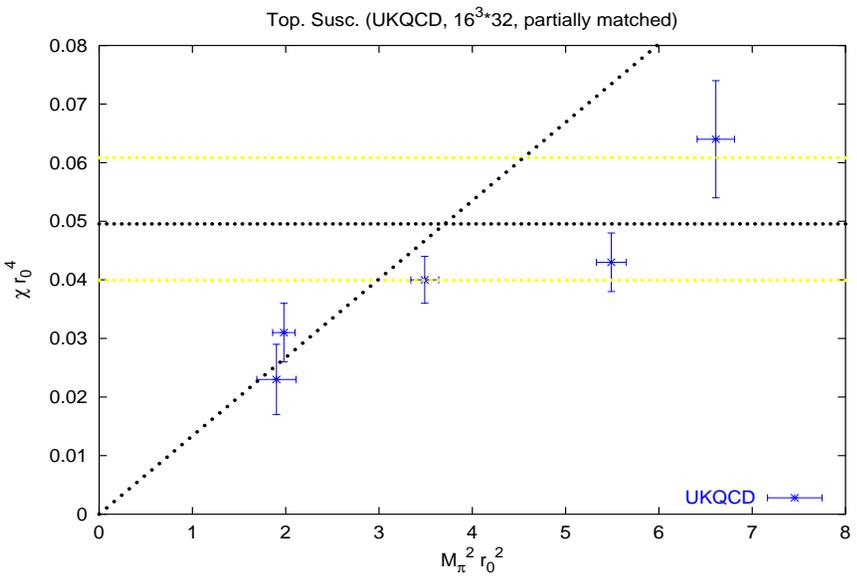,
height=7.7cm,width=11.7cm,angle=90}
\hfill
\epsfig{file=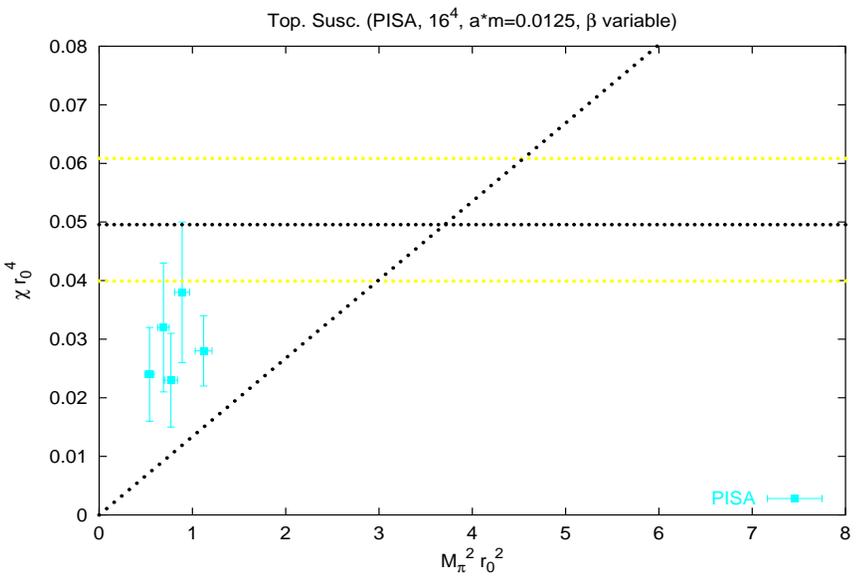,
height=7.7cm,width=11.7cm,angle=90}
\vspace{-2mm}
\\
\epsfig{file=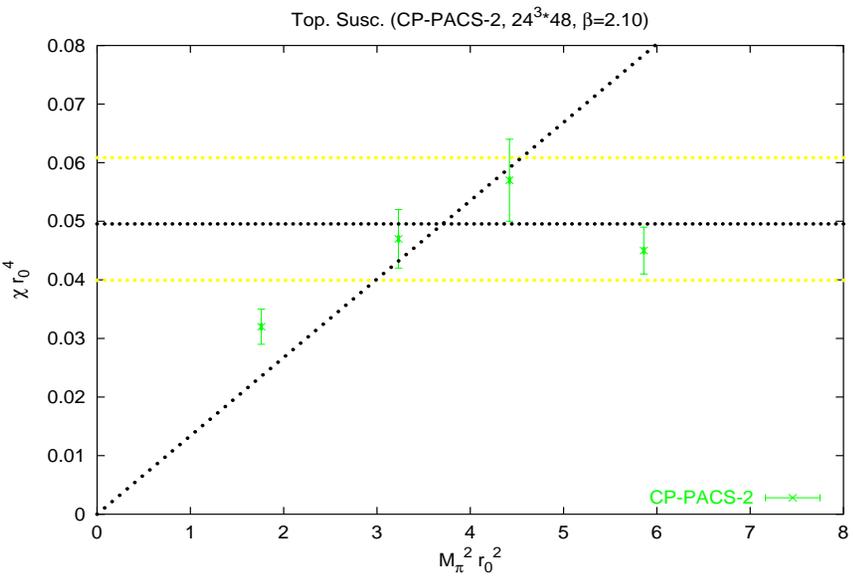,
height=7.7cm,width=11.7cm,angle=90}
\hfill
\epsfig{file=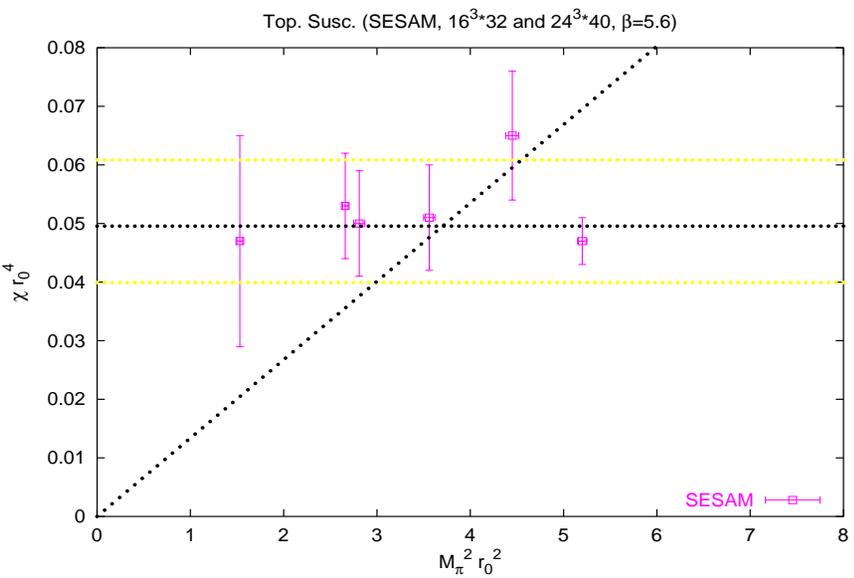,
height=7.7cm,width=11.7cm,angle=90}
\end{center}
\vspace{-8mm}
\caption{\sl\small
Data for the topological susceptibility versus the quark mass in $\Nf\!=\!2$
QCD by CP-PACS, UKQCD, SESAM and the Pisa group. The two asymptotic constraints
(based on $F_\pi\!=\!93\,\mr{MeV}$ and
$\ch_\infty\!=\!(190\!\pm\!10\,\mr{MeV})^4$, respectively) are shown; the data
are supposed to lie beneath either line, i.e.\ in the lower right sector (for
details see text). Data taken from \cite{CPPACS,UKQCD,SESAM,PISA,Aoki}.}
\end{figure}

From Fig.\ 4 one sees that the data generated by the CP-PACS collaboration on
the larger lattice ($24^3\!\times\!48$ at $\be\!=\!2.10$) as well as the
results obtained by UKQCD and the Pisa group are ``better'' in the sense that
they barely support an entirely flat behaviour, but they still seem to be at
the edge of being compatible with the constraints imposed by the known
asymptotic behaviour for small quark masses.
For the CP-PACS data at $\be\!=\!2.10$ and the UKQCD data the hypothesis that
the true 2-flavour topological susceptibility curve lies in the sector beneath
either one of the two indicated lines seems conceivable, in particular if one
takes into account that the horizontal line itself is subject to some
uncertainty%
\footnote{The dotted line assumes $\ch_\infty\!=\!(190\,\mr{MeV})^4$. Using
$\ch_\infty\!=\!(180\,\mr{MeV})^4$ or $\ch_\infty\!=\!(200\,\mr{MeV})^4$ shifts
the line to 0.04 or 0.06, respectively. Note that the situation for the chiral
constraint is different; here additional positive terms on the r.h.s.\ of the
GOR-relation $M_\pi^2 F_\pi^2\!=\!2\Sigma m+...$ would reduce its slope, and
the bound on $\Sigma$ is relatively tight, since the experimental determination
$F_\pi\!\simeq\!93\,\mr{MeV}$ is rather precise.}.
The remaining constraints --~the first derivative of $\ch(m)$ must always be
positive, the second always negative~-- seem to be obeyed with the
(unsignificant) exception of one point in each one of these two simulations.
Thanks to relatively large error bars, it seems that the data obtained by
the SESAM-collaboration are at the edge of being compatible with the two
asymptotic constraints.
For the data due to the Pisa group the situation looks less promising:
Each one of their data points is compatible, on a $2\si$-level, with the
linear constraint applicable in the deep chiral regime, but the combination
of the five points is not.
For obvious reasons nothing can be said about the first two derivatives from
either the SESAM or the Pisa data, the error bars are just too large.

\begin{figure}[t]%5
\vspace{-1mm}
\begin{center}
\epsfig{file=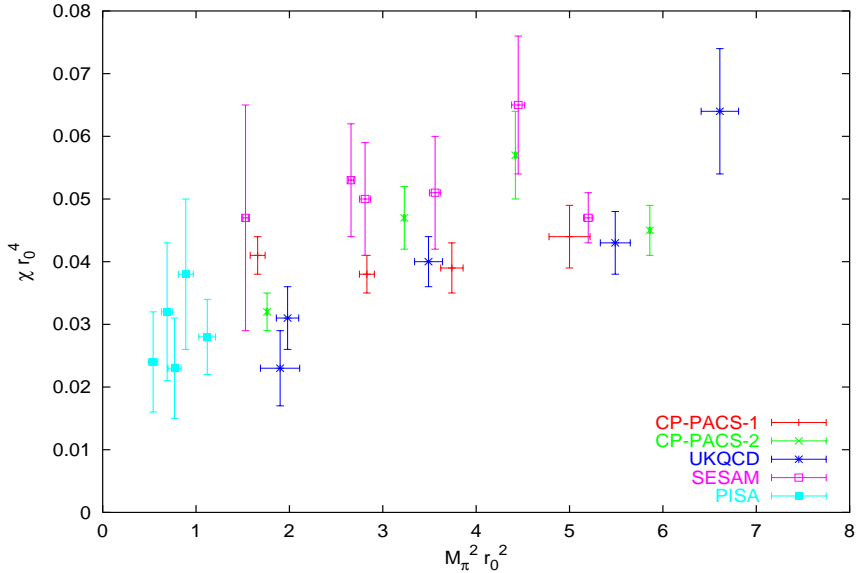,height=7.7cm,width=11.7cm,angle=0}
\end{center}
\vspace{-7mm}
\caption{\sl\small
Collection of recently published data for the topological susceptibility in
$\Nf\!=\!2$ QCD.}
\end{figure}

%The aim of the present paper is to clarify the situation regarding this point
%as far as possible by taking all available lattice measurement of the
%topological susceptibility in full QCD with 2 active flavours into account
%and sharpening theoretical expectations.
%Fig.\ 5 shows the combined data obtained by the CP-PACS collaboration
%\cite{CPPACS,Aoki}, UKQCD\cite{UKQCD,Aoki}, and the Pisa group \cite{PISA}.
Another simple cross-check is to combine the results obtained by the various
collaborations in a single plot and to see whether there is an obvious trend
for the data to depend on the group, an endeavour undertaken in Fig.\ 5.
The first impression is that the data are vaguely consistent with each other,
but even for the combined data it is not clear whether they show all of the
transient behaviour we expect to see with active flavours which are neither
light enough to produce a linear curve through the origin nor heavy enough to
have the topological susceptibility flatten out at the saturated quenched
level: They seem to indicate a {\em positive slope\/} (though the band is
rather fuzzy), but it is not clear whether the combined
data do exhibit the {\em negative curvature\/} genuine to the transition
region anticipated at $m\!=\!O(88\mr{MeV})$ or $M_\pi^2 r_0^2\!=\!O(3)$ for
$\Nf\!=\!2$ (cf.\ Fig.\ 1).

\begin{table}
\begin{center}
\begin{tabular}{|c|cccc|}
\hline
CP-PACS-1&
$16^3\times32$&$\be\!=\!1.95$&$c_{{}_\mr{SW}}\!\!=\!\!1.53$&
%$a_\rh^\mr{XL}\!=\!0.153\,\rm{fm}$
\\
\hline
$\ka$&0.1410&0.1400&0.1390&0.1375\\
$M_\pi/M_\rh$&0.586&0.688&0.751&0.805\\
\hline
$M_\pi^2 r_0^2$&1.66&2.83&3.74&5.00\\
$\ch r_0^4$&0.041&0.038&0.039&0.044\\
$a_\si[\mr{fm}]$&0.170&0.181&0.193&0.204\\
$V[\mr{fm}^4]$&109.&141.&182.&227.\\%V=V_\si
$x\!\equiv\!V\Sigma m$&42.0&92.1&157.&263.\\%x=x_\si
\hline
\end{tabular}
\end{center}
\vspace{-6mm}
\begin{center}
\begin{tabular}{|c|cccc|}
\hline
CP-PACS-2&
$24^3\times48$&$\be\!=\!2.10$&$c_{{}_\mr{SW}}\!\!=\!\!1.47$&
%$a_\rh^\mr{XL}\!=\!0.108\,\rm{fm}$
\\
\hline
$\ka$&0.1382&0.1374&0.1367&0.1357\\
$M_\pi/M_\rh$&0.575&0.690&0.757&0.806\\
\hline
$M_\pi^2 r_0^2$&1.76&3.23&4.44&5.86\\
$\ch r_0^4$&0.032&0.047&0.057&0.045\\
$a_\si[\mr{fm}]$&0.113&0.120&0.126&0.134\\
$V[\mr{fm}^4]$&107.&138.&167.&215.\\%V=V_\si
$x\!\equiv\!V\Sigma m$&43.7&103.&171.&292.\\%x=x_\si
\hline
\end{tabular}
\end{center}
\vspace{-6mm}
\begin{center}
\begin{tabular}{|c|ccccc|}
\hline
UKQCD&
$16^3\times32$&{\small partially}&{\small matched}&{}&{}
\\
\hline
$\be$&5.20&5.20&5.20&5.26&5.29\\
$c_{{}_\mr{SW}}$&2.02&2.02&2.02&1.95&1.92\\
$\ka$&0.13565&0.1355&0.1350&0.1345&0.1340\\
$M_\pi/M_\rh$&0.567&0.600&0.688&0.791&0.830\\
\hline
$M_\pi^2 r_0^2$&1.90&1.98&3.49&5.49&6.61\\
$\ch r_0^4$&0.023&0.031&0.040&0.043&0.064\\
$a_{r_0}[\mr{fm}]$&0.094&0.097&0.103&0.104&0.102\\
$V[\mr{fm}^4]$&10.3&11.7&14.8&15.4&14.1\\%V=V_{r_0}
$x\!\equiv\!V\Sigma m$&9.1&10.7&23.9&39.1&43.1\\%x=x_{r_0}
\hline
\end{tabular}
\end{center}
\vspace{-6mm}
\begin{center}
\begin{tabular}{|c|cccc|cc|}
\hline
SESAM&
$16^3\times32$&$\be\!=\!5.6$&$c_{{}_\mr{SW}}\!\!=\!\!0$&{}&$24^3\times40$&dito
\\
\hline
$\ka$&0.1575&0.1570&0.1565&0.1560&0.1580&0.1575\\
$M_\pi/M_\rh$&0.692&0.763&0.813&0.834&0.574&0.704\\
\hline
$M_\pi^2 r_0^2$&2.81&3.56&4.45&5.20&1.53&2.66\\
$\ch r_0^4$&0.050&0.051&0.065&0.047&0.047&0.053\\
$a_{r_0}[\mr{fm}]$&0.084&0.091&0.095&0.098&0.080&0.085\\
$V[\mr{fm}^4]$&6.5&9.1&10.6&12.1&23.1&28.9\\%V=V_{r_0}
$x\!\equiv\!V\Sigma m$&8.5&15.1&21.8&29.1&16.4&35.6\\%x=x_{r_0}
\hline
\end{tabular}
\end{center}
\vspace{-6mm}
\begin{center}
\begin{tabular}{|c|ccccc|}
\hline
PISA&
$16^4$&$am\;=\!$&$\!\!0.0125$&($N_{\!f}\!=\!2$&stag.)
\\
\hline
$\be$&5.40&5.50&5.55&5.60&5.70\\
%$M_\pi/M_\rh$&?.???&?.???&?.???&?.???&?.???\\
\hline
$M_\pi^2 r_0^2$&0.54&0.69&0.77&0.89&1.12\\
$\ch r_0^4$&0.024&0.032&0.023&0.038&0.028\\
%$a_m[\mr{fm}]$&0.176&0.137&0.123&0.107&0.085\\
%$V[\mr{fm}^4]$&63.1&23.1&15.2&8.66&3.43\\%V=V_m
%$x\!\equiv\!V\Sigma m$&15.8&7.4&5.4&3.6&1.8\\%x=x_m
$a_\si[\mr{fm}]$&0.174&0.134&0.125&0.108&0.086\\
$V[\mr{fm}^4]$&60.3&21.3&15.9&8.78&3.55\\%V=V_\si
$x\!\equiv\!V\Sigma m$&15.1&6.8&5.7&3.6&1.8\\%m=m_si
\hline
\end{tabular}
\end{center}
Tables~1-5:~{\small\sl Key data of the five studies.
For error bars see Figs.\ 1-5 and Refs.\ \cite{CPPACS,UKQCD,SESAM,PISA,Aoki}.
In all cases the Leutwyler-Smilga parameter $x\!\equiv\!V\Sigma m$ has been
estimated by re-expressing it, through the GOR relation, as
$x\!=\!V M_\pi^2 F_\pi^2/2\!=\!V(M_\pi^2 r_0^2)(F_\pi^2/2r_0^2)$ with
$F_\pi\!\simeq\!93\,\mr{MeV}, r_0\!\simeq\!0.49\,\mr{fm}$.}
\end{table}

The consistency between the various data sets cannot be discussed, however,
without taking into account that the groups have adopted rather different
simulation strategies.
CP-PACS has decided to simulate at fixed $\be$, going more chiral by tuning
$\ka$ closer to $\ka_\mr{crit}$.
As discussed above, this approach {\em reduces\/} the physical lattice spacing
$a$ for low $M_\pi/M_\rh$.
In order to keep $a$ constant, one would have to reduce $\be$ appropriately
while increasing $\ka$, an approach which --~if successfully implemented~--
produces ``matched ensembles''.
This is the philosophy adopted by UKQCD.
For technical reasons ($c_{{}_\mr{SW}}$ is not known nonperturbatively for
$\be\!<\!5.2$) this strategy had to be sacrificed for their two most chiral
points, i.e.\ their data set is, strictly speaking, just ``partially matched''.
SESAM, on the other hand, has decided to take everywhere the simplest choice,
i.e.\ to vary $\ka$ at fixed $\be$ while not making any attempt to improve the
Wilson fermions.
An orthogonal strategy has been adopted by the Pisa group: They simulate
(using the staggered rather than a Wilson-type formulation for their dynamical
flavours) at fixed $am$, thus decreasing $\be$ in order to go more chiral.
Note that --~contrary to what happens in a series of fixed-$\be$ simulations~--
the Pisa approach makes the lattice {\em coarser\/} as one goes more chiral.

A possible inconsistency between the results by the different groups
%--~which, despite some assertions \cite{UKQCD},
%%controversy / polemics / far reaching statements / claims
%seems not obvious given the current size of the error bars~--
could come from a variety of reasons:

(a) Scaling violations: Because of the computational costs involved in full
QCD simulations the latter are typically performed on relatively coarse
lattices (cf.\ Tables 1-5).
For the simulations analyzed in this article scaling violations are supposed to
be ``small'' since $O(a)$ effects have been removed by mean-field techniques
(CP-PACS) or non-perturbatively (UKQCD) or they might be numerically suppressed
due to a small $a$ (SESAM) or absent anyway (Pisa group).
Nontheless, the physical lattice spacing varying from one data point to another
in the same study might imply unequal shifts due to cutoff effects.

(b) Chirality violation with Wilson-type fermions: Wilson fermions (SESAM)
violate the chiral symmetry at nonzero lattice spacing, and this holds true,
too, for all ultralocally improved descendents (e.g.\ the clover-fermions used
by CP-PACS and UKQCD).

(c) Finite-volume effects: From Tables 1-5 one learns that the five studies
have been conducted in boxes of rather different physical four-volume $V$
and at rather different values of the Leutwyler-Smilga parameter
$x\!\equiv\!V\Sigma m$.
Hence, if finite-volume effects are not fully under control, potential
inconsistencies of type (a) or (b) might get masked by them.

I have a few comments on each one of these points:

On (a): UKQCD claimed that in order to get simulation results which indicate a
quark mass dependence of the topological susceptibility it was {\em crucial\/}
to use the strategy of generating matched ensembles, rather than running at
fixed $\be$ \cite{UKQCD}.
Despite this assertion it holds true that in principle either method should
work: If one could do both a continuum and a chiral extrapolation, it should
not matter at all where in the $(\be,\ka)$ plane the data were collected.
In practice a continuum extrapolation at fixed quark mass is unaffordable at
the present time, and hence each group has to live with the single data point
for a given value of $M_\pi^2r_0^2$.
Since both UKQCD and CP-PACS work with $O(a)$ improved actions for which a
continuum extrapolation was shown to bring just a minor shift in the case of
pseudoscalar masses \cite{Aoki}, it is natural to expect the topological
susceptibility data to be reasonably close to their continuum values too.
If unequal magnitudes of scaling violations in a fixed-$\be$ strategy were
indeed responsible for the majority of the CP-PACS-1 ``flatness problem'', then
one would expect the data generated by CP-PACS at $\be\!=\!1.95$ to be
subject to larger shifts for {\em large\/} quark masses (where the lattice is
coarser) versus more trustworthy at {\em small\/} quark masses (where $a$ is
smaller), but from Fig.\ 3 one gets the impression that the problem with that
particular set is primarily in its most chiral point.
Scaling violations may indeed account for the bulk of the difference between
the UKQCD and CP-PACS-1 data sets, but it seems that such a difference is much
more likely associated with the respective {\em overall differences between
typical lattice spacings\/} (Tables 1 and 3 indicate that UKQCD works on much
finer lattices than those employed by CP-PACS at $\be\!=\!1.95$) rather than
whether these sets are matched or not.
This is also the conclusion one is led to when comparing the CP-PACS-1 data
(Table 1) to those entitled CP-PACS-2 (Table 2): The latter seem to indicate a
quark mass dependence even though they were generated at fixed $\be\!=\!2.10$.
Since the volume parameters $V$ and $x$ are in one-to-one analogy between
CP-PACS-1 and CP-PACS-2 (see Tables 1 and 2), the only difference is the
lattice spacing which is significantly smaller at the higher $\be$-value.

On (b): The quark mass dependence of the topological susceptibility in the
extreme chiral regime is derived from the flavour singlet axial WT-identity%
\footnote{See e.g.\ Ref.\ \cite{Aoki}. It may also be derived from chiral
perturbation theory, since the latter is just a neat way of implementing the
constraints imposed by the axial WT-identity.}.
With Wilson quarks (or descendents which do not satisfy the Ginsparg-Wilson
relation) this identity receives additional terms \cite{RossiTesta}.
This may result in the lattice topological susceptibility $\hat\ch$ being
enhanced compared to its continuum counterpart $\ch$.
It should be mentioned, however, that this effect is, in principle, accounted
for by the additive renormalization term in (\ref{chirenadd}).
Hence, on a practical level, an accurate determination of $M\!=\!M(\be,\Nf,m)$
should be sufficient to deal with this challenge.

On (c): As indicated in footnote 2, for the topological susceptibility to raise
linearly as a function of the (small) quark mass it is important that the
volume is ``infinite'', i.e.\ in more technical terms that the Leutwyler-Smilga
parameter (\ref{xdef}) is large.
How ``large'' it has to be for this behaviour to show up and what pattern the
topological susceptibility follows if this criterion is not satisfied will be
discussed in the following two sections.
As we shall see, finite-volume effects are the only source of contamination
of the lattice topological susceptibility for which an a-priory
{\em quantitative assessment\/} may be achieved.

%%%%%%%%%%%%%%%%%%%%%%%%%%%%%%%%%%%%%%%%%%%%%%%%%%%%%%%%%%%%%%%%%%%%%%%%%%%%%%%

\section{Leutwyler-Smilga regimes}

Considering a theory which exhibits the spontaneous breakdown of a continuous
global symmetry (e.g.\ of the $SU(\Nf)_A$ group for QCD) in a finite volume,
one expects to see both symmetry restoration phenomena and the onset of
Goldstone boson production, if the box is taken sufficiently ``small'' or
``large'', respectively.
An obvious question is what sets the scale, i.e.\ by which standards does the
box have to be small or large to trigger one or the other type of phenomena~?

The naive guess is the lightest particle at hand, i.e.\ the mass $M_\pi$ of the
pseudo-Goldstone boson produced in the infinite-volume limit.
From this one would expect symmetry restoration to take place for
$M_\pi L\!\ll\!1$ and SSB to become manifest for $M_\pi L\!\gg\!1$.

As Leutwyler and Smilga (LS) have shown \cite{LeutwylerSmilga}, there is, in
addition, the parameter
\beq
x \equiv V\Sigma m
\label{xdef}
\eeq
which also depends on the quark mass and which proves particularly useful in
a lattice context, since it decides which one of the limits $m\!\to\!0,
V\!\to\!\infty$ is in the ``inner'' position and therefore winning%
\footnote{In QCD the two limits $\lim_{m\to0}$, $\lim_{V\to\infty}$ are known
not to commute: The chiral condensate vanishes if $m\!\to\!0$ is performed
first, but it tends to $-\Sigma$, if the volume is sent to infinity first
and then the chiral limit is taken. In a numerical analysis the two limits
cannot be considered separately; the upshot is that lattice data will reflect
the former situation for $x\!\ll\!1$ and the latter for $x\!\gg\!1$.}.
In other words: For $x\!\ll\!1$ the chiral symmetry is almost restored, i.e.\
quarks and gluons are the dominant degrees of freedom, whereas for $x\!\gg\!1$
the chiral symmetry is (though the box volume is still finite) effectively
broken, i.e.\ long range Green's functions are dominated by Goldstone
excitations.
Furthermore, the value of $x$ has a bearing on whether standard%
\footnote{The meaning is: an observable not related to the
$U(1)_A$-issue, e.g.\ $M_\pi,F_\pi,M_\rh,V_{q \bar q}$, but not $M_{\et'}$.}
physical observables depend on the net topological charge $q$ of the gauge
background or not \cite{LeutwylerSmilga}:
In the small LS-regime ($x\!\ll\!1$) all observables depend massively
on $|q|$, and the finite-volume partition function is entirely dominated by the
contribution from the charge zero sector \cite{LeutwylerSmilga}, viz.\
\beq
Z_\nu \simeq \de_{\nu\,0}
\qquad\qquad(x\!\ll\!1)
\;.
\label{LSpeak}
\eeq
On the other hand, in the large LS-regime ($x\!\gg\!1$) standard observables
prove basically independent of $|q|$ (for a numerical check and some
qualifications see \cite{Durr:2001ei}), and the finite-volume partition
function gets broad and gaussian \cite{LeutwylerSmilga}, viz.\
\beq
Z_\nu \simeq {1\ovr\sqrt{2\pi\si^2}}\exp\{-{\nu^2\ovr2\si^2}\}
\quad,\qquad
\si^2={V\Sigma m\ovr\Nf}
\qquad\qquad(x\!\gg\!1)%eigentlich:(x\!\gg\!1,M_\piL\!\ll\!1)
\;.
\label{LSlarge}
\eeq

It is important to note that the LS-parameter $x$ and the ``naive'' parameter
\beq
y \equiv M_\pi L
\label{ydef}
\eeq
are not the same: In either one of the regimes with pronounced ($x\!\gg\!1$) or
mild ($x\!\simeq\!1$) SSB the box may be large ($y\!\gg\!1$), intermediate
($y\!\simeq\!1$) or even small ($y\!\ll\!1$) w.r.t.\ the pion correlation
length classification, whereas in the symmetry restoration regime ($x\!\ll\!1$)
such a distinction does not make sense, since there the pion is not a useful
degree of freedom.
That $x$ and $y$ cover different aspects of the box being ``small'' or ``large''
is also borne out by the fact that the minimum box length scales differently as
a function of the inverse quark mass, if one wants the box to be large w.r.t.\
either criterion, i.e.
\bea
L_\mr{\min}\propto(1/m)^{1/4}&\mr{for}&m\to0\;\mr{at}\;\mr{fixed}\;
(\mr{large})\;x\\
L_\mr{\min}\propto(1/m)^{1/2}&\mr{for}&m\to0\;\mr{at}\;\mr{fixed}\;
(\mr{large})\;y,
\label{scaling}
\eea
which simply follows from the definition (\ref{xdef}) of $x$ and the GOR
relation, respectively.
Furthermore, it is worth noticing that in a lattice context the
LS-classification refers to the mass of the {\em sea\/}-quarks (i.e.\ the
dynamical quarks which influence the weight of a gauge configuration through
the determinant), whereas the conventional one refers to the mass of the
{\em current\/}-quarks (i.e.\ those from which hadronic observables are built).

In QCD the chiral symmetry restoration transition is intricately related to
the deconfinement transition, and this is the reason why even the combination
of the parameters $x$ and $y$ is not sufficient to describe the broad
properties of the finite-volume system.
This is easily seen from the fact that the confinement-deconfinement transition
takes place at a critical temperature $T_c\!\simeq\!275\,\mr{MeV}$ in pure
$SU(3)$-gluodynamics and at a somewhat reduced temperature
$T_c\!\simeq\!175\,\mr{MeV}$ in QCD with $\Nf\!=\!2$ massless flavours.
In other words: $T_c$ stays at the same order of magnitude while both $x$
(cf.\ (\ref{xdef})) and $y$ (cf.\ \ref{ydef})) decrease from infinity to zero
as the quarks are tuned from infinitely heavy to weightless.
Hence a third scale must be involved which --~contrary to $x$ and $y$~-- barely
depends on $\Nf$ and $m$.
It is tempting%
\footnote{This is not intended to overrule the more traditional view that the
transition occurs when the hadrons are packed so densely that they start to
overlap, though it does have the decisive advantage over the latter that it
does not try to predict the transition from the properties of one phase only.
(The true condition is of course the equalness of the free energy densities.)
In addition, it should be stated that no claim is made that instantons are
truly responsible for the confinement-deconfinement transition (at least random
superpositions of instantons do not reproduce the string tension
\cite{Chen:1999ct}), but that --~because of the relationship the chiral and the
confinement transition have in QCD~-- a complete assessment of whether the box
is ``small'' or ``large'' needs a scale which does {\em not depend on the quark
masses\/}. The scale $L_\mr{inst}$ fulfills this requirement and turns out to
be numerically in the right order, and that is all we need for the present
analysis.}
to identify the missing scale with the one due to {\em instantons\/},
i.e.\ to introduce the parameter
\beq
z \equiv {L_\mr{min} \ovr L_\mr{inst}}
\label{zdef}
\eeq
where $L_\mr{min}$ is the smallest%
\footnote{In addition, bosonic/fermionic fields are supposed to be
periodic/antiperiodic in that direction.}
linear dimension of the box, and $L_\mr{inst}$ is a typical linear size
associated to an instanton, e.g.
$L_\mr{inst}\!=\!2\rh_\mr{inst}\!\simeq\!0.7\,\mr{fm}$ or
$L_\mr{inst}\!=\!V_\mr{inst}^{1/4}\!\simeq\!1\,\mr{fm}$ \cite{SchaferShuryak},
since such a scale is, to a large extent, independent of the number and mass of
the active flavours.
For $z\!\ll\!1$ the system is in the deconfined state, i.e.\ quarks and gluons
represent the appropriate degrees of freedom, whereas for $z\!\gg\!1$
confinement is manifest, i.e.\ asymptotic states are colour singlets, and the
associate composite fields are the building blocks of the effective description
appropriate in that regime.

As we shall see in the following section, the subtle interplay between the
LS-scale $x$ (which depends on the quark masses) and the instanton size (which
barely does) is crucial for an understanding of the overall behaviour of the
topological susceptibility curve in QCD.

%%%%%%%%%%%%%%%%%%%%%%%%%%%%%%%%%%%%%%%%%%%%%%%%%%%%%%%%%%%%%%%%%%%%%%%%%%%%%%%

\section{Finite-volume partition function with ``granularity''}

We are now in a position to exploit the fact that one may actually
{\em derive\/} the topological susceptibility from the finite-volume partition
function, and for the latter several versions are available in the literature.

\bigskip

In the large LS-regime the situation is particularly simple, since from the
representation (\ref{LSlarge}) of the QCD finite-volume partition function one
gets $\<\nu^2\>\!=\!V\Sigma m/\Nf$ and hence the linearly raising behaviour
(\ref{susclarge}) of the topological susceptibility, valid for $x\!\gg\!1$.
Analogously, one may compute $\ch$ from $Z_\nu$ in the other two regimes, since
Leutwyler and Smilga give, in their landmark paper \cite{LeutwylerSmilga}, also
the generalization of (\ref{LSlarge}) to arbitrary $x$:
\beq
Z_\nu^\mr{XPT}=
\left|
\begin{array}{cccc}
I_{\nu}(x)&I_{\nu+1}(x)&\cdots&I_{\nu+\Nf-1}(x)\\
I_{\nu-1}(x)&I_{\nu}(x)&\cdots&I_{\nu+\Nf-2}(x)\\
\vdots&\vdots&\ddots&\vdots\\
I_{\nu-\Nf+1}(x)&I_{\nu-\Nf+2}(x)&\cdots&I_{\nu}(x)
\end{array}
\right|
\;.
\label{LSall}
\eeq
It is easy to check that, upon using the asymptotic form
$I_\nu(x)\!\simeq\!e^{x}(1-(\nu^2\!-\!1)/(8x))/\sqrt{2\pi x}$ for $x\!\gg\!1$
of the modified Bessel function of integer order, the representation
(\ref{LSall}) of the partition function reduces to (\ref{LSlarge}) for
$x\!\gg\!1$, hence reproducing the linearly raising form (\ref{susclarge}) of
$\ch$ in the large LS-regime.
Furthermore, from the representation $I_\nu(x)\!\simeq\!x^{|\nu|}/(2^{|\nu|}
\Gamma(|\nu|\!+\!1))$ appropriate in the opposite case $x\!\ll\!1$ one gets the
asymptotic behaviour \cite{LeutwylerSmilga}
\beq
\ch \propto m^{\Nf}
\qquad\qquad(x\!\ll\!1)
\label{suscsmall}
\eeq
characteristic for the small LS-regime.
For an $x$ value which is neither small nor large compared to one, a numerical
evaluation of $\<\nu^2\>\!=\!\Sigma \nu^2Z_\nu^\mr{XPT}/\Sigma Z_\nu^\mr{XPT}$
is needed, and the result may be compared to the asymptotic behaviour
(\ref{susclarge}) or (\ref{suscsmall}) for $x\!>\!1$ or $x\!<\!1$.
One of these comparisons is demonstrated on the l.h.s.\ of Fig.\ 6.
There are two important lessons we can learn from that graph.
First, {\em finite-volume effects have a decreasing effect\/} on the
topological susceptibility.
Second, the topological susceptibility as derived from (\ref{LSall}) and the
large-$x$ asymptotic form (\ref{susclarge}) are {\em reasonably close for
moderate $x$-values\/} already.
To be definite: For $V\!=\!10.125\,\mr{fm}^4$ and
$\Sigma\!=\!(288\,\mr{MeV})^3$, the mark $x\!=\!1$ is reached at a quark mass
$m\!=\!6.26\,\mr{MeV}$, and there the finite-$x$ value is about 50\% smaller
than the asymptotic prediction.
Taking the quark mass 10 times larger, $m\!=\!62.6\,\mr{MeV}$
(i.e.\ $x\!=\!10$), the finite-$x$ value lies less than 10\% below the
asymptotic (large-$x$) prediction (cf.\ Fig.\ 6).

\begin{figure}[t]%6
\epsfig{file=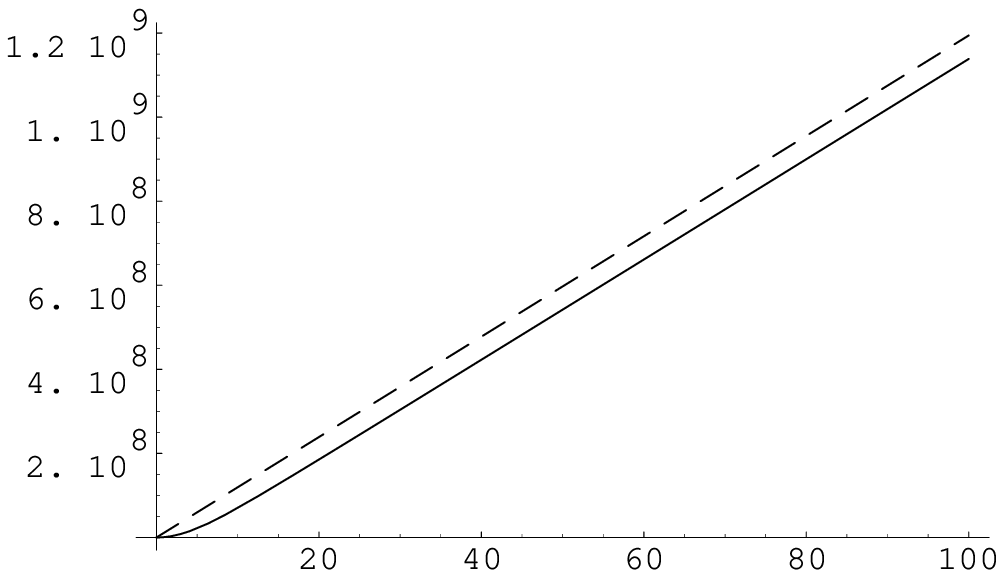,
height=5cm,width=7cm,angle=0}
\hfill
\epsfig{file=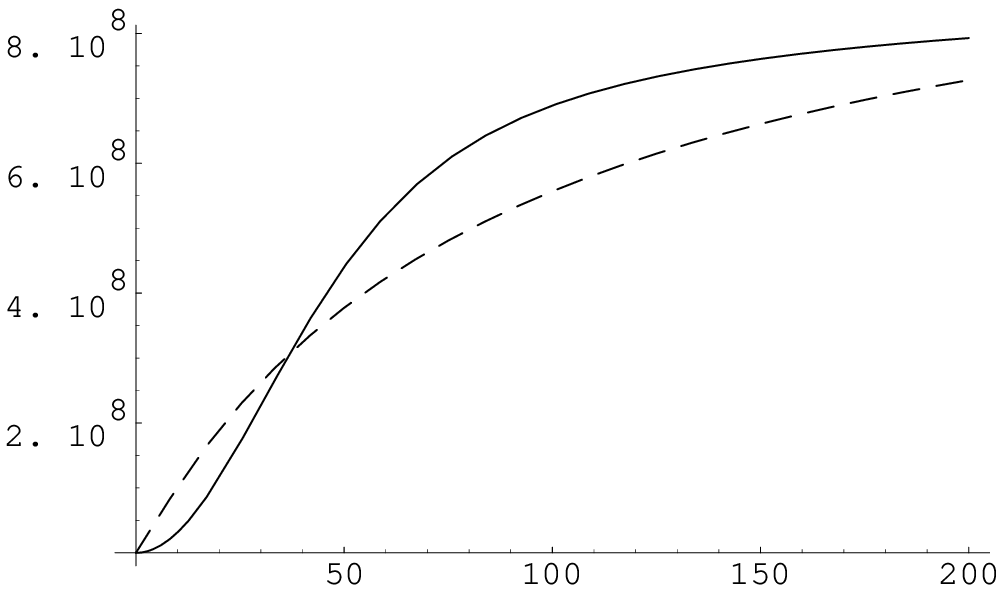,
height=5cm,width=7cm,angle=0}
\vspace{-3mm}
\caption{\sl\small
LHS: Topological susceptibility (without instanton effects) [in $\mr{MeV}^4$]
versus quark mass [in $\mr{MeV}$] as derived from (\ref{LSall}) (full curve),
and the large-$x$ asymptotic form (\ref{susclarge}) (dashed line) in a box of
$V\!=\!2\,(1.5\,\mr{fm})^4$ with $\Sigma\!=\!(288\,\mr{MeV})^3$. Since
$\Nf\!=\!2$, the small-$x$ asymptotic form is a parabola (cf.\
(\ref{suscsmall})). RHS: Illustration of the scenario in which the topological
susceptibility in a finite volume (full) ``overshoots'' the infinite-volume
curve (dashed) for some intermediate quark masses, hence explaining the CP-PACS
``flatness problem'' showing up in\ Fig.\ 3. In this illustration, the
``granularity'' of the QCD vacuum, as discussed in the text, is already taken
into account. This ``overshooting'' scenario is excluded, since finite volume
effects may only have a decreasing effect on $\ch$ (cf.\ LHS and the
``granular'' construction (\ref{Zcombi}) discussed in the text).}
\end{figure}

The first statement is important, because it rules out a possible explanation
of the ``flatness problem'' in the CP-PACS determination of the topological
susceptibility on the smaller ($16^3\!\times\!32, \be\!=\!1.95$) lattice,
apparent in Fig.\ 3: Knowing that the topological susceptibility drops, in the
small LS-regime, with a power-like behaviour -- cf.\ (\ref{suscsmall}) -- and
thus stays massively below the linear curve (\ref{susclarge}), one might have
suspected that $\ch$ could ``overshoot'' and lie above the linear curve in an
intermediate regime.
Such a hypothetical behaviour, as illustrated in the r.h.s.\ of Fig.\ 6, which
would provide a simple explanation why the leftmost point in Fig.\ 3 is so
high, is definitely ruled out by the graph on the l.h.s.\ of Fig.\ 6; in other
words, the ``chiral constraint'' indicated in Figs.\ 3 and 4 is to be taken
seriously.

The second statement is important, because it implies that in a lattice study
the minimum box length needed to avoid finite-volume effects in the topological
susceptibility is much smaller than what one might have suspected by analogy
with the standard pion correlation length criterion.
The standard criterion says that in the broken phase the minimum box length
$L_\mr{min}$ has to be larger than $\xi_\pi$ by up to a factor 4 to avoid
finite-volume effects in hadronic Green's functions.
If the same factor (per linear dimension) would be relevant for the
LS-parameter $x$, the latter would have to be of order $4^4\!=\!256$, which,
as one can see from Tables 1-5, is hardly the case in any of the simulations
discussed in this article.
Hence from the l.h.s.\ of Fig.\ 6 one takes the good news that the actual value
needed is {\em much smaller\/}; already $x\!\geq\!10$ seems sufficient to make
the curve lie reasonably close to the large-$x$ line.
Furthermore, comparing $\ch$ as predicted by the QCD partition functions
(\ref{LSlarge}) and (\ref{LSall}) offers the potential benefit to quantify
finite-volume effects in the topological susceptibility.

\bigskip

The next step is to recall that there is an important limitation to either
topological susceptibility curve shown in the l.h.s.\ of Fig.\ 6, i.e.\ this
limitation is inherent to both the solid curve (numerically obtained from
(\ref{LSall})) and the dashed line (derived from (\ref{LSlarge})):
Either function {\em grows without bound\/}, i.e.\ they both violate the other
asymptotic constraint, $\ch\to\ch_\infty$ for $m\to\infty$, as discussed in
Sec.\ 2 and illustrated in Fig.\ 1.

In such a situation it is crucial to reflect on the physical basis of the
second constraint, valid at large quark masses.
It is easy to see that the relevant physics is due to {\em instantons\/};
they are sufficient to get an understanding of the numerical value of the
quenched topological susceptibility, $\ch_\infty$.
Assume the total four-volume $V$ of the box to be sufficiently large, so
that it can be divided into $N$ hypercubic boxes of volume $V_0$ each, and
--~without loss of generality~-- $N$ even.
In the Instanton Liquid Model (ILM) the typical volume occupied by an
instanton or anti-instanton is $V_\mr{inst}\!\simeq\!1\,\mr{fm}^4$
\cite{SchaferShuryak}, so we expect $V_0$ to be roughly of that size.
Assume now that each one of these boxes is populated, {\em at random\/}, by
either an instanton or an anti-instanton.
What is the resulting topological susceptibility ?
The answer is easy: The topological charge distribution is a modified binomial
distribution (note that $\nu/2\in\mr{\bf Z}$)
\beq
Z^\mr{ILM}_\nu=
\left(
{N \atop N/2+\nu/2}
\right)
\label{PFinstbinomial}
\eeq
which, after it has been smoothed to fill the odd-$\nu$ entries too, may be
approximated by
\beq
Z_\nu^\mr{ILM} \simeq {1\ovr\sqrt{2\pi\si^2}}\exp\{-{\nu^2\ovr2\si^2}\}
\quad,\qquad
\si^2=N={V\ovr V_0}
\qquad\qquad
(N\!\gg\!1)
\;.
\label{PFinstgaussian}
\eeq
This distribution, in turn, implies $\<\nu^2\>\!\simeq\!V/V_0$, and henceforth
\beq
\ch_\infty \simeq {1\ovr V_0}
\;,
\label{suscbox}
\eeq
which means that, if the cell volume $V_0$ was identical to
$V_\mr{inst}\!=\!1\,\mr{fm}^4$, the quenched
topological susceptibility would evaluate to $1.52\,10^9\,\mr{MeV}^4$.
Hence, all we need to do, in order to get a more standard value of the quenched
topological susceptibility, is to increase the size of the little boxes, i.e.\
to pack the instantons a little bit looser:
\beq
V_0 \simeq 1.44\,\mr{fm}^4 \simeq (1.10\,\mr{fm})^4
\quad\Longleftrightarrow\quad
\ch_\infty \simeq 1.05\,10^9\,\mr{MeV}^4 \simeq (180\,\mr{MeV})^4
\;.
\label{adjust}
\eeq
The basic lesson from this exercise is that it is possible to get a rough
quantitative understanding of the quenched topological susceptibility from
a simple statistics consideration involving nothing but the parameter
$V_0$, the average volume occupied by an instanton, hence {\em omitting all the
dynamics between instantons\/} which is incorporated in the ILM
\cite{SchaferShuryak}.

\bigskip

\begin{figure}[t]%7
\epsfig{file=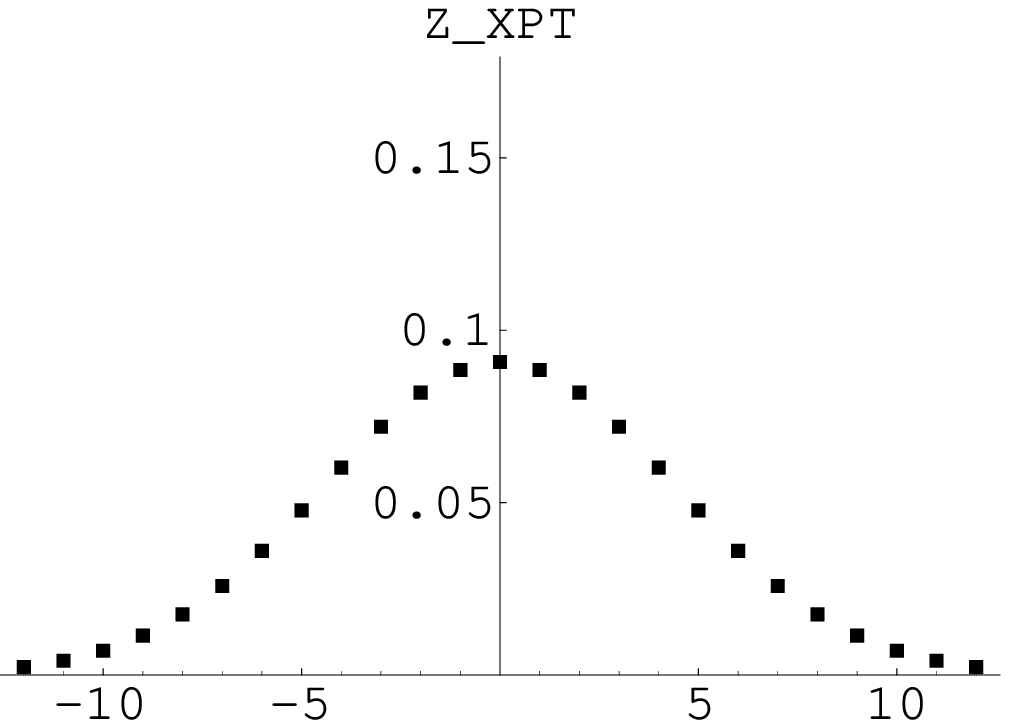,
height=5cm,width=7cm,angle=0}
\hfill
\epsfig{file=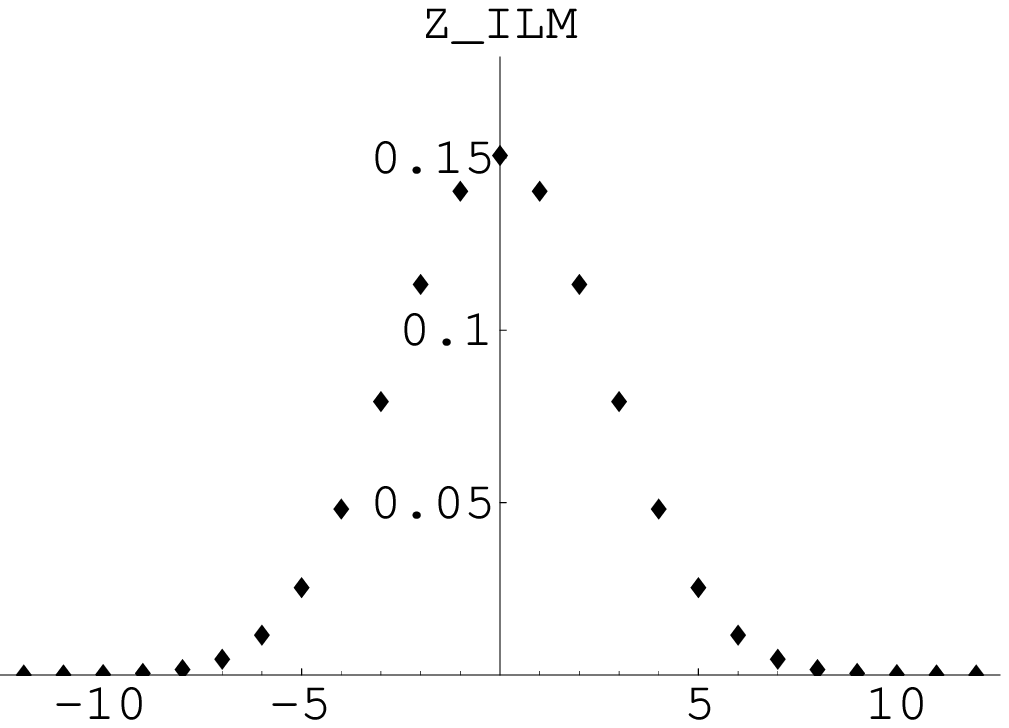,
height=5cm,width=7cm,angle=0}
\vspace{3mm}
\\
\epsfig{file=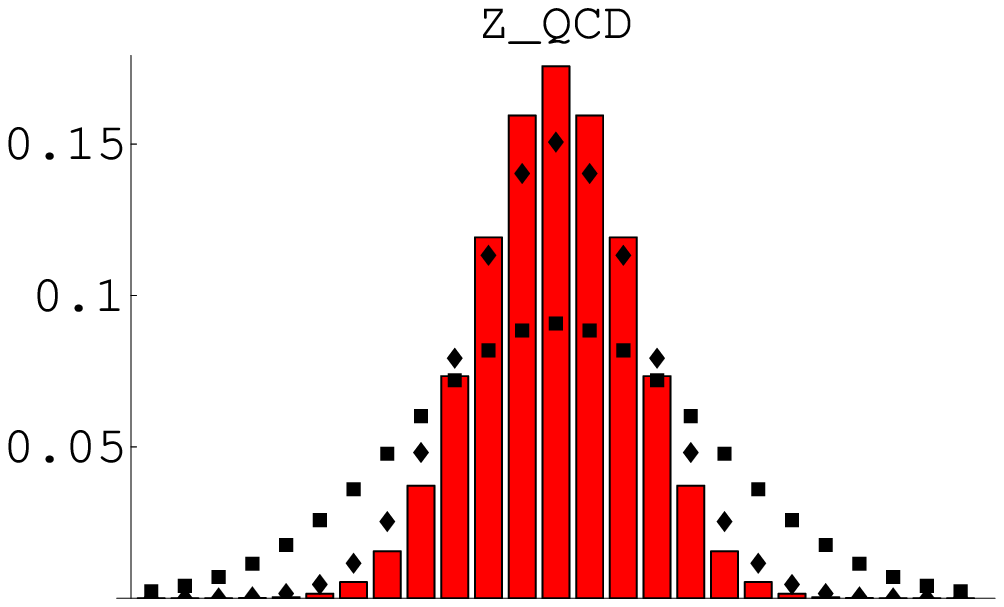,
height=5cm,width=7cm,angle=0}
\hfill
\epsfig{file=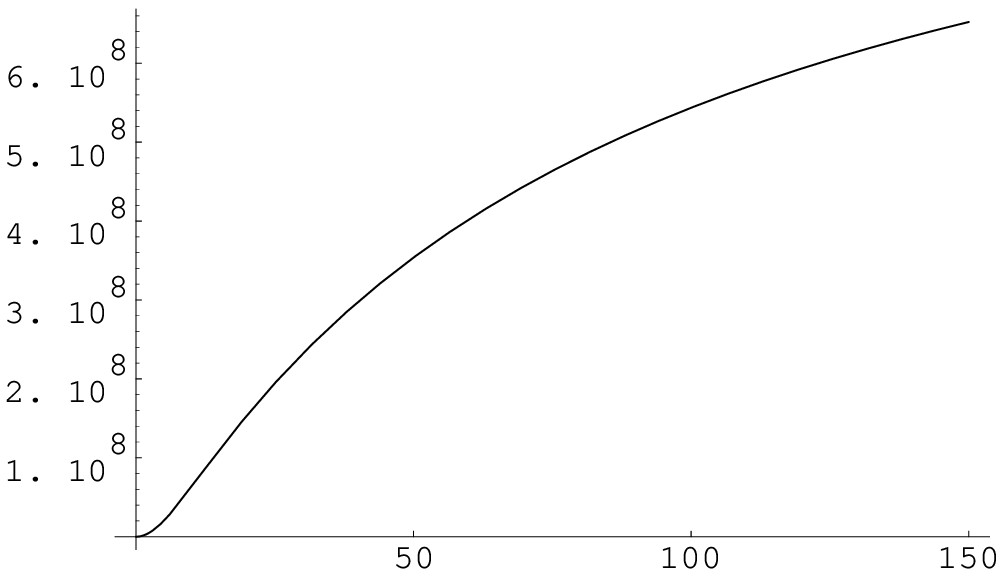,
height=5cm,width=7cm,angle=0}
\vspace{0mm}
\caption{\sl\small
Normalized QCD finite-volume partition function $Z_\nu$ as predicted by the
Leutwyler-Smilga form (\ref{LSall}) (top left) and by the instanton inspired
form (\ref{PFinstgaussian}) (top right) for $V\!=\!2\,(1.5\,\mr{fm})^4$ and
$m\!=\!250\,\mr{MeV}$. Following the XPT prediction, the topological sectors
with $|\nu|\!\geq\!8$ should be populated, but the ILM statistics consideration
demands that they are heavily suppressed, since instantons cannot be
compressed. The construction (\ref{Zcombi}) assumes that the two high-$|\nu|$
suppression mechanisms do not interfere. The resulting ``QCD finite-volume
partition function with granularity'' is narrower than either one of its
predecessors (bottom left), and the topological susceptibility curve derived
from it (bottom right) shows all essential features: power-like behaviour for
$x\!<\!1$, almost linear behaviour for $x\!>\!1$ and
$\Sigma m/\Nf\!<\!\ch_\infty$, saturation for
$\Sigma m/\Nf\!\geq\!\ch_\infty$.}
\end{figure}

It is now easy see why the topological susceptibility as derived from the
finite-volume partition function (\ref{LSall}) is, in general, too large.
To this end it is useful to consider the distribution of topological charges
as predicted by (\ref{LSall}) in a box volume typical for full QCD simulations,
say $V\!=\!2\,(1.5\,\mr{fm})^4\!=\!10.125\,\mr{fm}^4$, at a quark mass of, say,
$250\,\mr{MeV}$.
As one can see from Fig.\ 7, in this regime the partition function $Z^\mr{XPT}$
as given in (\ref{LSall}) is {\em wider\/} than the partition function
$Z^\mr{ILM}$ as given in (\ref{PFinstgaussian}).
In particular, for sufficiently high $|\nu|$ the former indicates a sizable
population of that topological sector, whereas the latter is almost zero.
This means that the (leading order) chiral prediction neglects the basic fact
that instantons form an {\em incompressible fluid\/} \cite{SchaferShuryak},
i.e.\ it is impossible to pack such a large number of instantons into the box
as the tail of the $Z^\mr{XPT}$ distribution suggests.
This calls for an improvement on $Z^\mr{XPT}$ which incorporates the unique 
feature of $Z^\mr{ILM}$ that instantons occupy a fixed volume.
In the absence of an exact analytic solution it is natural to assume that the
two suppression mechanisms (captured in (\ref{LSall}, \ref{PFinstgaussian}),
respectively) operate {\em independent of each other\/}, i.e.\ to construct
the QCD finite-volume partition function by just multiplying%
\footnote{It is worth mentioning that this is reasonable regardless of what are
the respective energy and entropy components in the two partition functions.
For a single degree of freedom (\ref{Zcombi}) means that the energy differences
w.r.t.\ $E_0$ add up. For a larger number of degrees of freedom either
partition function is, in general, the product of a Boltzmann-type energy
suppression and an entropy factor. In our case $Z_\nu^\mr{XPT}$ is likely of
this ``mixed'' type (even though the axial WT-identity from which it takes its
origin does not disclose any information on this point), but the second factor
$Z_\nu^\mr{ILM}$ as constructed in (\ref{PFinstgaussian}) is pure entropy.}
\footnote{Note that the multiplicative rule (\ref{Zcombi}) w.r.t.\ $\nu$ is
equivalent to a folding prescription in the basis w.r.t.\ the dual variable
$\th$: The likelyhood that the system is found in the interval
$[\th_\mr{min},\th_\mr{max}]$ is $Z^\mr{XPT}(\th') \cdot Z^\mr{ILM}(\th'')$
integrated over all $\th',\th''$ with the constraint that the sum $\th'+\th''$
is in the interval $[\th_\mr{min},\th_\mr{max}]$.}
the two asymptotic%
\footnote{The meaning is: appropriate, by themselves, for $m\to0$ or
$m\to\infty$, respectively. The choice of the representation (\ref{LSall}) for
$Z_\nu^\mr{XPT}$ and of (\ref{PFinstgaussian}) for $Z_\nu^\mr{ILM}$ makes sure
that (\ref{Zcombi}) is appropriate for arbitrary $x$ and any $V$ large enough
to accommodate several instantons (e.g.\ $V\!\geq10\,\mr{fm}^4$).}
partition functions
\beq
Z^\mr{QCD}_\nu = Z^\mr{XPT}_\nu \cdot Z^\mr{ILM}_\nu
\;,
\label{Zcombi}
\eeq
as is illustrated in the lower left plot in Fig.\ 7.
As one can see, the resulting distribution is narrower than either one of the
asymptotic distributions it was constructed from.
This still holds true in the opposite regime of small quark masses where the
relationship is reversed (there $Z^\mr{XPT}$ is narrower than $Z^\mr{ILM}$
which is quark mass independent).
As a result of this the topological susceptibility is {\em smaller\/} than both
the XPT-prediction and the ILM-inspired value.
This is particularly easy to see in the limit of large box volumes; there
either partition function entering the r.h.s.\ of (\ref{Zcombi}) tends to be
gaussian (with variance $\si^2\!=\!V\Sigma m/\Nf$ and $\si^2\!=\!V\ch_\infty$,
respectively), and the multiplication prescription (\ref{Zcombi}) then implies
\beq
\ch(\Nf,m)=1/(\Nf/\Sigma m+1/\ch_\infty)
\;,
\label{suscombi}
\eeq
which indeed fulfills $\ch(\Nf,m)\!\leq\!\min(\Sigma m/\Nf,\ch_\infty)$.
Note that (\ref{suscombi}) is nothing but the member $n\!=\!1$ out of the
phenomenological family (\ref{chipheno}).
In addition, two points should be mentioned:
First: The nice feature that finite-volume effects in $\ch$ disappear for
moderate $(x,V)$ combinations already (say for $x\!\geq\!10,
V\!\geq\!10\,\mr{fm}^4$) has survived.
This follows from the construction (\ref{Zcombi}) of the ``granular'' QCD
finite-volume partition function, but one may also verify it by carefully
comparing the last graph in Fig.\ 7 to the dashed (``infinite volume'') curve
on the r.h.s.\ of Fig.\ 6 -- the graph in Fig.\ 7 was generated for
$V\!=\!2\,(1.5\,\mr{fm})^4\!=\!10.125\,\mr{fm}^4$, hence reaching $x\!=\!10$
at $m\!\simeq\!62.6\,\mr{MeV}$.
Second: As a specific finite-volume effect the topological susceptibility
curve may show, in a certain (typically narrow) window of (typically small)
quark masses, a positive curvature -- a feature absent in the infinite-volume
limit.

\bigskip

We shall conclude this paragraph with a short reflection on what we have
achieved with the ``granular'' construction (\ref{Zcombi}) of the (true) QCD
finite-volume partition function.

On the conceptual level a specific observable, the topological susceptibility
$\ch$, has been linked to a more general concept, the QCD finite-volume
partition function with granularity.
The name reminds one that the expression (\ref{LSall}) which resulted from a
careful evaluation, at finite $x$, of the QCD partition function in terms of
the chiral Lagrangian to order $O(p^2)$ \cite{LeutwylerSmilga} got modified to
account for the ``granular structure'' of the QCD vacuum, due to instantons.
Note that the ``granular'' partition function depends separately on $V$ and $m$
-- not only on the product of them, as (\ref{LSlarge}) and (\ref{LSall}) did.
The construction is not ad hoc, because the distribution $Z_\nu^\mr{QCD}$
predicts, besides $\ch$, another observable, namely the mass dependence of the
sectoral chiral condensate in a finite volume, hence promising an interesting
generalization of the results by Damgaard \cite{Damgaard:1999ij}.
On the other hand, it is clear that the construction is neither a model nor
exact, since subleading effects (where the dynamics of QCD enters) have been
neglected; the instanton effect taken into account is {\em pure entropy\/}%
\footnote{In this respect it should be noted that the suffix ``ILM'' is,
strictly speaking, a misuse of the name, since the real instanton liquid model
does account for the dynamics between instantons \cite{SchaferShuryak}.}.

On the practical level, the construction (\ref{Zcombi}) has managed to select,
for purely theoretical reasons, out of the phenomenological family
(\ref{chipheno}) the specific member (\ref{suscombi}) as the one which is, in
the limit of large box volumes, most promising.
This provides us with an overall fitting curve with just two parameters which
is supposed to be reasonable for arbitrary quark mass values.
Furthermore, at any given box volume a detailed comparison of the topological
susceptibility curve as predicted by (\ref{Zcombi}) to the ``infinite volume''
form (\ref{suscombi}) allows one to come up with a {\em quantitative assessment
of finite-volume effects\/} on $\ch$.
This may prove useful in future simulations of full QCD, since it enables one
to choose the box volume just as large as needed in order to keep finite-volume
effects tolerable, hence providing some help in the challenging task to
allocate computational resources as reasonably as possible.
The limiting factor is, of course, that even the demonstrated smallness (within
the accuracy of the construction (\ref{Zcombi})) of finite-volume effects in
$\ch$ is no proof that finite-volume effects in other interesting observables
would be small.

%%%%%%%%%%%%%%%%%%%%%%%%%%%%%%%%%%%%%%%%%%%%%%%%%%%%%%%%%%%%%%%%%%%%%%%%%%%%%%%

\section{Fits, Cuts and Speculations}

We are now ready for a detailed analysis of the available data for the
topological susceptibility in 2-flavour QCD.

The most valuable output of the theoretical considerations presented above is
a well-motivated functional form against which we may fit the data {\em over
the entire range of quark masses\/}.
This is particularly useful because, as mentioned in the second section, the
bulk of the data is in the transition regime; hence functional forms which are
inspired by an expansion appropriate in one asymptotic regime are not guaranteed
to yield reliable results for the physical parameters we are interested in.

For large enough box volumes (for details see below) formula (\ref{suscombi})
will be applicable.
It contains two parameters, $\Sigma$ and $\ch_\infty$, on which the simulations
will disclose some information.
Regarding the chiral condensate in the chiral limit, it is important to note
that the way it will be determined (via fitting (\ref{suscombi}) to the lattice
data) is {\em logically independent\/} from the usual approach of measuring
the condensate at various quark masses via the trace of the Green's function of
the Dirac operator and then extrapolating to zero mass.
The procedure to determine $\Sigma$ from the topological susceptibility curve
does not require a subtraction or difficult-to-know renormalization constants
that arise in more conventional approaches.

\begin{table}[b]
\begin{center}
\begin{tabular}{|c|cccc|cc|cc|}
\hline
&$\!$CP-PACS$\!$&$\!$UKQCD$\!$&$\!$SESAM$\!$&$\!$PISA$\!$&3-avg&4-avg&all&sel\\
\hline
$2\Nf/(F_\pi^2 r_0^2)$&24.6&46.4&16.4& 5.6&29.1&23.3&15.8&24.1\\
$F_\pi$ [MeV]         &162.&118.&199.&340.&149.&167.&203.&164.\\
$\Sigma^{1/3}$ [MeV]  &418.&338.&478.&684.&395.&425.&484.&420.\\
\hline
$\ch_\infty r_0^4$&0.063&0.083&0.076&0.035&0.074&0.064&0.054&0.063\\
$\ch_\infty$ [$10^9$\,MeV${}^4$]&1.7&2.2&2.0&0.9&1.9&1.7&1.4&1.7\\
$\ch_\infty^{1/4}$ [MeV]&202.&216.&211.&174.&210.&203.&194.&202.\\
\hline
\end{tabular}
\end{center}
\vspace{-1mm}
Table~6:~{\small\sl
Results of the fits of the data by the individual groups (Fig.\ 8) or of the
combination of all or a selection of them (Fig.\ 9) to the functional form
(\ref{suscombi}) and of $1/(\ch r_0^4)$ as a function of $1/(M_\pi^2 r_0^2)$ to
a straight line, weighting the results 2:1. For both the inverse chiral
symmetry breaking parameter (upper part) and the quenched topological
susceptibility (lower part) the average over the first 3 or 4 collaborative
results is taken in the first line. Successive lines are derived using
$r_0\!=\!0.49\,\mr{fm}$ and, to convert $F_\pi$ into $\Sigma$, the values for
$M_\pi, m$ quoted in footnote 5.}
\end{table}

Fig.\ 8 displays the data by CP-PACS (on the $24^3\!\times\!48$ lattice only),
UKQCD, SESAM and the Pisa group together with individual interpolating curves
which represent both reasonable fits of the data to the functional form
(\ref{suscombi}) and of the inverse data versus the inverse argument to a
straight line.
The respective values for $\Sigma$ and $\ch_\infty$ are tabulated in Table 6
(along with further details on how they are determined).
A naive average (performed in the top line in either part of the table) yields
$\Sigma\!\simeq\!(425\,\mr{MeV})^3$ and
$\ch_\infty\!\simeq\!(203\,\mr{MeV})^4$, respectively, or
$\Sigma\!\simeq\!(395\,\mr{MeV})^3$ and
$\ch_\infty\!\simeq\!(210\,\mr{MeV})^4$ if the values suggested by the Pisa
group study are not included.
A direct fit to the combined data (see Fig.\ 9) suggests
$\Sigma\!\simeq\!(484\,\mr{MeV})^3$ and
$\ch_\infty\!\simeq\!(194\,\mr{MeV})^4$.
It seems that the outcome of this combined analysis is influenced, to a huge
extent, by the Pisa data -- even stronger than the naive average.

\begin{figure}[p]%8
\vspace{-4mm}
\begin{center}
\epsfig{file=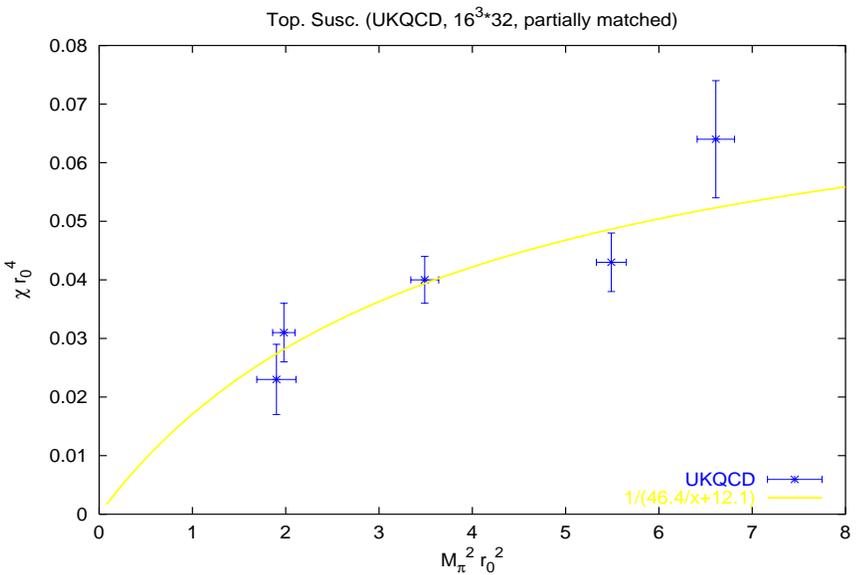,
height=7.7cm,width=11.7cm,angle=90}
\hfill
\epsfig{file=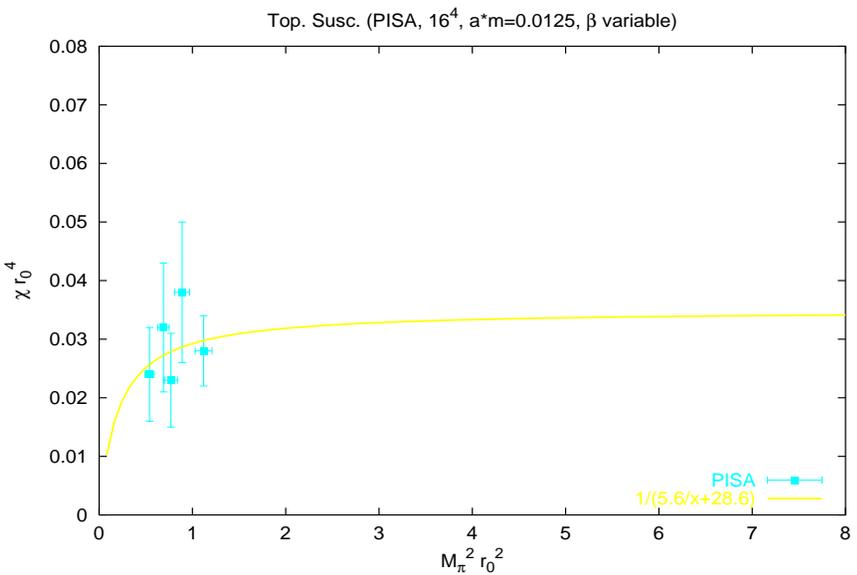,
height=7.7cm,width=11.7cm,angle=90}
\vspace{-2mm}
\\
\epsfig{file=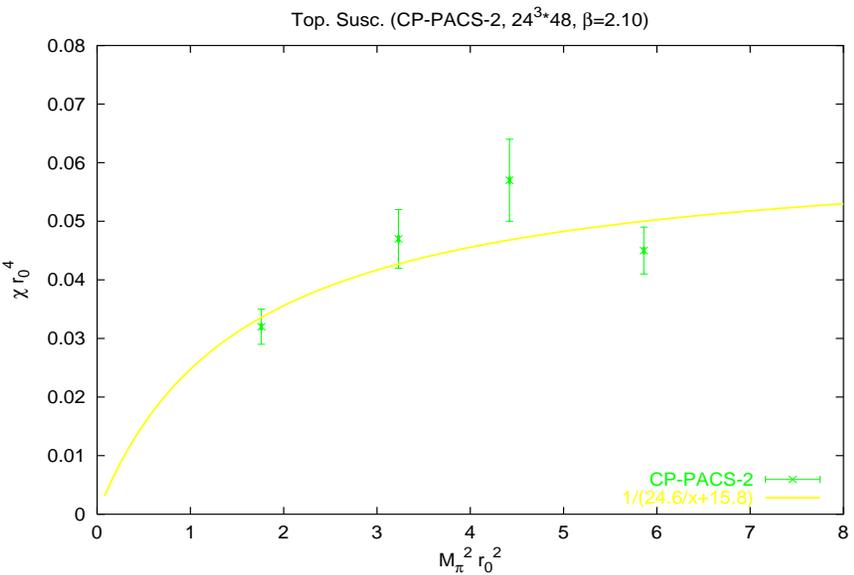,
height=7.7cm,width=11.7cm,angle=90}
\hfill
\epsfig{file=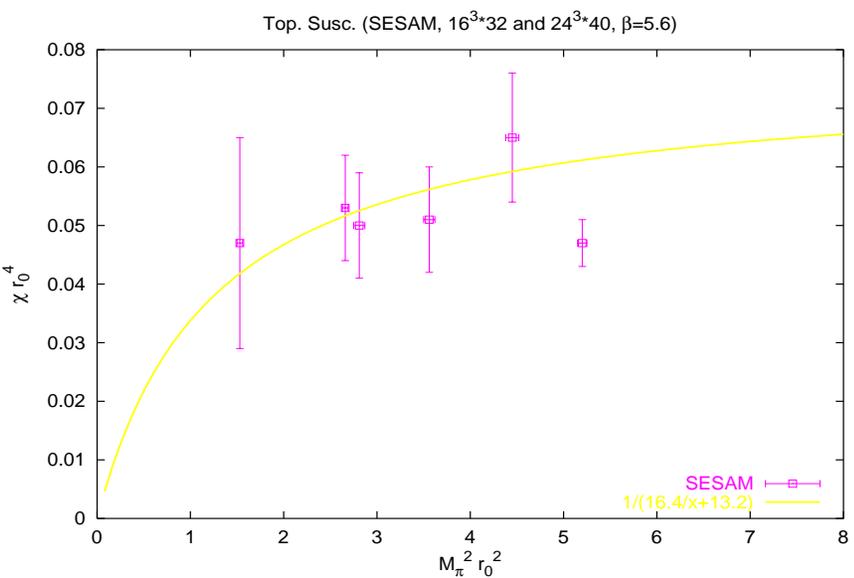,
height=7.7cm,width=11.7cm,angle=90}
\end{center}
\vspace{-8mm}
\caption{\sl\small
Data generated by CP-PACS, UKQCD, SESAM and the Pisa group for the topological
susceptibility versus the quark mass in 2-flavour QCD together with individual
fits to the functional form (\ref{suscombi}). The slopes at zero and the
asymptotic values at $m\!\to\!\infty$ determine $\Sigma$ and $\ch_\infty$,
respectively. Data taken from \cite{CPPACS,UKQCD,SESAM,PISA,Aoki}.}
\end{figure}

\begin{figure}[t]%9
\vspace{-2mm}
\begin{center}
\epsfig{file=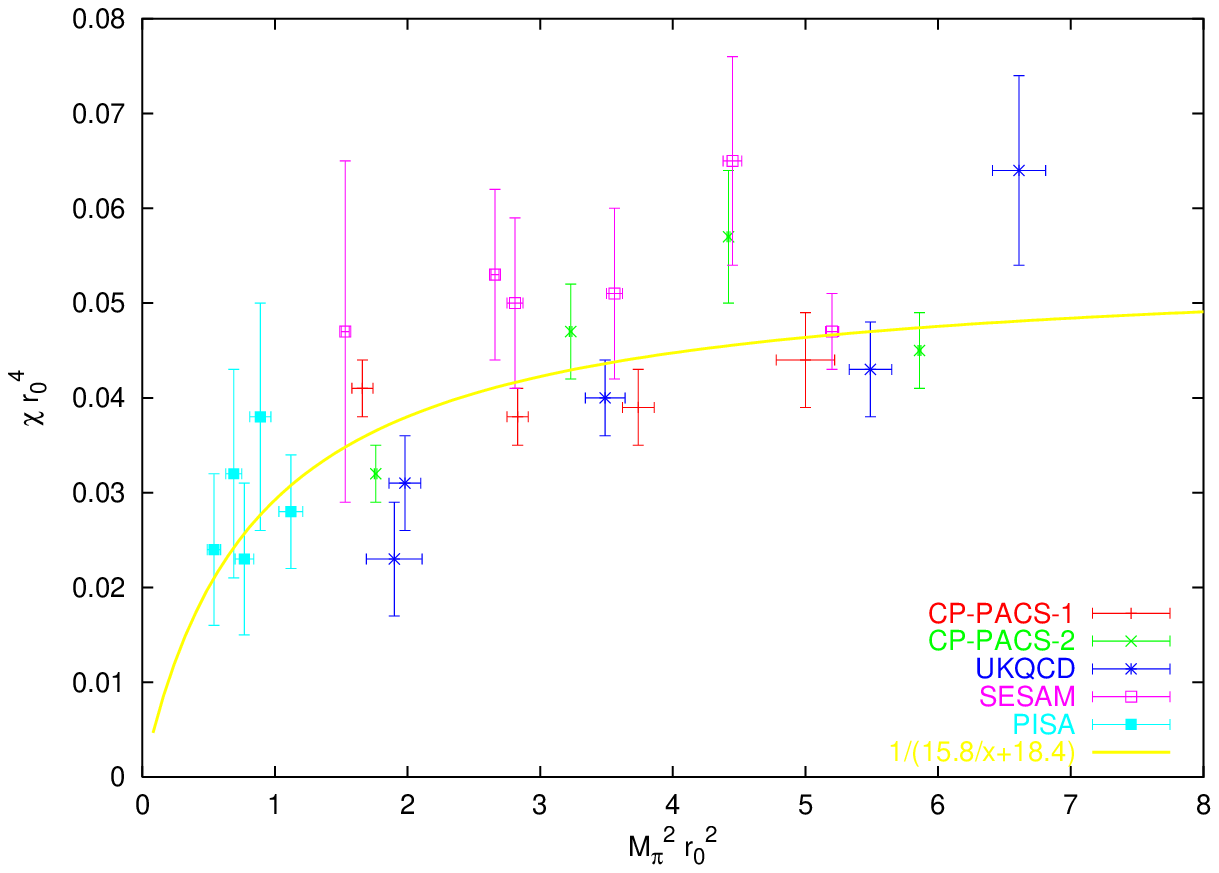,
height=7.7cm,width=11.7cm,angle=0}
\\
\epsfig{file=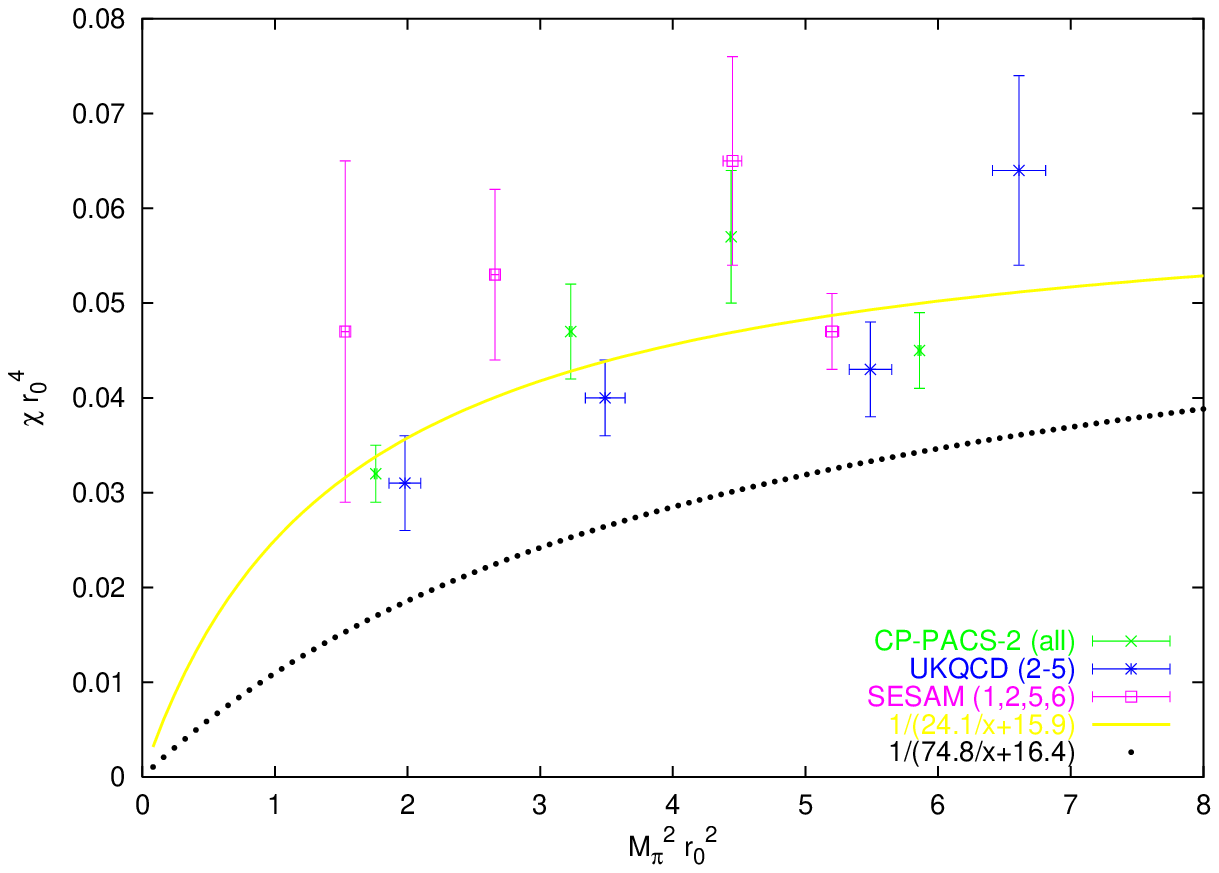,
height=7.7cm,width=11.7cm,angle=0}
\end{center}
\vspace{-7mm}
\caption{\sl\small
Fitting all data (top) or only those which pass the cuts (\ref{cuts}) (bottom)
to the functional form (\ref{suscombi}). Either way $\Sigma$ seems to be much
larger than expected, as one can see from comparing the fitted curve to the
one anticipated --~based on $\Sigma\!=\!(288\,\mr{MeV})^3$ and the most
optimistic value $\ch_\infty\!=\!(200\,\mr{MeV})^4$~-- which is included in the
second plot. Data taken from \cite{CPPACS,UKQCD,SESAM,PISA,Aoki}.}
\end{figure}

To ensure the quality of the final phenomenological parameters it seems
necessary to require the data points to meet certain criteria to be included
in the analysis.
What we have learned from the previous two sections is that we should not
only require $a$ to be reasonably small to keep lattice artefacts under
control, but we should also demand the LS-parameter $x$ and the volume $V$ to
be reasonably large to keep finite-size effects on $\ch$ moderate and to be
able to accommodate a reasonable number of instantons.
What I dare proposing, for these reasons, is to apply the following
{\em cuts\/} on the data
\beq
(\mr{a})\;\;a\!\leq\!0.15\,\mr{fm}\quad,\qquad
(\mr{b})\;\;x\!\geq\!10\quad,\qquad
(\mr{c})\;\;V\!\geq\!10\,\mr{fm}^4\;,
\label{cuts}
\eeq
irrespective of the circumstances%
\footnote{Cut (a) is unfair w.r.t.\ the CP-PACS data; it does not care whether
the gluons are improved or not.}.
From Tables 1-5 one sees that this battery of cuts eliminates the CP-PACS data
on the smaller grid and those by the Pisa group.
In addition, the UKQCD data set looses its most chiral point, and for the
SESAM data only those generated on the larger grid and the two least chiral
ones on the smaller grid survive.
Table 6 contains, in its last column, the results of a direct analysis of the
combined (``selected'') data which have survived the cuts (\ref{cuts}).
The net outcome almost coincides with the naive average of the results by the
individual groups; in the light of the sizable deviations%
\footnote{This is the reason why no error bars are given in Tables 1-6. It
seems that overall systematic effects dominate over statistical fluctuations.}
between the results by the various groups (cf.\ Table 6) I quote
\bea
\Sigma=(390...430\,\mr{MeV})^3\quad&\mr{or}&\quad F_\pi=145...170\,\mr{MeV}
\label{fpifinal}
\\
\ch_\infty&=&(200...210\,\mr{MeV})^4
\label{chifinal}
\eea
as the current result for $\Sigma$ and $\ch_\infty$ when determined via the
topological susceptibility curve (and without continuum extrapolation) in
2-flavour QCD.

For the quenched topological susceptibility the result (\ref{chifinal}) is
slightly larger than $\ch_\infty$ as determined in previous direct studies in
the $SU(3)$ theory \cite{TOPOLOGY}.
This is not disturbing, since the generally accepted value
$\ch_\infty\!\simeq\!(190...200\,\mr{MeV})^4$ is after a continuum
extrapolation, whereas the value (\ref{chifinal}) is gained at finite $a$;
hence the difference may --~as it is ``small''~-- be attributed to the $O(a^6)$
term showing up on the r.h.s.\ of eqn.\ (\ref{chirenadd}).

For the chiral symmetry breaking parameter the situation seems less pleasing;
the result (\ref{fpifinal}) is {\em substantially larger\/} than the generally
accepted $F_\pi\!\simeq\!93\,\mr{MeV}, \Sigma\!\simeq\!(288\,\mr{MeV})^3$, as
determined experimentally or via the GOR-relation, respectively
(cf.\ footnote 5).
This excess may either be due to lattice artefacts or it may reflect an
interesting $\Nf$-dependence of QCD close to the chiral limit.
We shall discuss these two possibilities in due course.

\bigskip

Regarding lattice artefacts as a possible reason why (\ref{fpifinal}) exceeds
our expectations, little can be said beyond the points already raised in
section 4.
The issue is not whether lattice artefacts are present --~they are, since
no continuum extrapolation has been attempted~-- but rather whether they are
numerically under control, i.e.\ whether the final result (\ref{fpifinal}) is
largely influenced by discretization effects or not.
The point is that we have just seen that the value (\ref{chifinal}) for
$\ch_\infty$ happened to be in rough agreement with our expectations even
{\em without a continuum extrapolation\/}, and if lattice artefacts are
``small'' for the quenched topological susceptibility, it is natural to expect
their influence on the other measured quantity, $\Sigma$ (or $F_\pi$) to be
tolerable too, and this is apparently not the case.

Lattice artefacts have been categorized, in section 4, as (a) scaling
violations and (b) chirality violation effects.
While these two types of lattice artefacts may, in general, be independent of
each other, with Wilson fermions and its descendents (as used in the
simulations discussed here) they go together, i.e.\ they jointly disappear in
the continuum-limit and only then.
Since in all the data which have passed the cuts (\ref{cuts}) smaller quark
masses go, in tendency, together with smaller lattice spacings (except for the
UKQCD data which are partially matched, cf.\ Table 3), we would expect the
deviations from the dotted curve (which reflects our expectations based on
(\ref{suscombi}) together with $\Sigma\!=\!(288\,\mr{MeV})^3,
\ch_\infty\!=\!(200\,\mr{MeV})^4$) to become smaller for low quark masses, if
(a) or (b) (or both of them) represent the source of the supposed excess.
The bottom part of Fig.\ 9 (where finite-volume effects are negligible) may
contain a hint in this direction: If we suppress the points by the SESAM
collaboration (which have huge error bars anyway) the deviation, by absolute
standards, of the data from the dotted curve seems to increase towards the
r.h.s.\ (where $a$ is, in tendency, larger).
In addition, it looks like there is a slight tendency for the UKQCD data to
lie beneath the CP-PACS-2 data.
This nicely corresponds with the order of typical lattice spacings as used in
these simulations (cf.\ Tables 2-5), even though unequal efforts have been
spent on improving the action.
These observations certainly support the view of lattice artefacts as the
primary reason why the value (\ref{fpifinal}) for the chiral order parameter
exceeds our expectations by roughly a factor 2 (which means a factor $\sqrt{2}$
for $F_\pi$), but they are far from conclusive.

\bigskip

The second possibility is that the values (\ref{fpifinal}, \ref{chifinal})
are indeed correct, in other words that the problem is not with the lattice
determinations of $\Sigma$ (or $F_\pi$) and $\ch_\infty$, but rather with our
expectations -- specifically with our knowledge regarding the chiral order
parameter.
This might come as a surprise, since it is generally believed hat the value
$F_\pi\!\simeq\!93\,\mr{MeV}$ or $\Sigma\!\simeq\!(288\,\mr{MeV})^3$ is rather
well known experimentally or through the GOR-relation, respectively.
The point is that the quoted (largely different) values need not necessarily be
in conflict with each other, since (\ref{fpifinal}) is determined in
$\Nf\!=\!2$ QCD, while the experimental value is in QCD with%
\footnote{This is not to be confused with the fact that the (non-strange)
$M_\pi$ would hardly change if the strange quark would be made infinitely
heavy. The statement is that $F_\pi$ might feel the influence of the strange
{\em sea\/}-quarks, hence spoiling the generally believed ``stand alone''
quality of the sector with 2 light flavours.}
$\Nf\!=\!3$.
The statement is that there might be an $\Nf$-dependence in
the topological susceptibility near the chiral limit {\em beyond the one
indicated in\/} (\ref{susclarge}).
In other words, there might be an $\Nf$-dependence in the chiral condensate
itself which makes $\Sigma_2$ as determined in the lattice studies under
discussion differ from the experimentally ``observed'' $\Sigma_3$, contrary to
the standard belief that $\Sigma_{\Nf}$ is%
\footnote{The understanding is: for the range of flavours we are considering
here, i.e.\ for $\Nf\!=\!2...4$.}
{\em largely independent\/} of $\Nf$.

The idea that $\Sigma$ might depend on $\Nf$ is neither new nor exotic.
It is known that QCD has a ``conformal window'' with restored chiral symmetry
if the number $\Nf$ of massless or light flavours is appropriately chosen, i.e.\
$N_{\!f\,\mr{crit}}'\!<\!\Nf\!<\!N_{\!f\,\mr{crit}}''$ \cite{Banks:1982nn}.
While the upper critical number for this phase is the number of flavours where
QCD looses asymptotic freedom, i.e.\ $N_{\!f\,\mr{crit}}''\!=\!33/2$ (for
$N_{\!c}\!=\!3$) \cite{Gross:1973id}, the lower critical number is much under
debate.
Originally, the Orsay group argued that $N_{\!f\,\mr{crit}}'$ could be so low
(say $N_{\!f\,\mr{crit}}'\!\simeq\!O(3)$, as opposed to the more standard view
$N_{\!f\,\mr{crit}}'\!\simeq\!O(7)$ \cite{Iwasaki:1998qv}) that the chiral
condensate in ordinary 3-flavour QCD might be marginal or even vanishing, hence
necessitating the extension of the chiral Lagrangian approach \cite{SXPTrev}
to the framework%
\footnote{With $\Nf$ close to $N_{\!f\,\mr{crit}}'$ the supposed order
parameter $\Sigma$ becomes small, and the ``correction terms'' on the r.h.s.\
of the GOR-relation $M_\pi^2 F_\pi^2\!=\!2\Sigma m+...$ turn larger than the
``leading'' term. As a result of this, the chiral expansion must be reordered
-- at the price of reducing the predictivity of the theory at a given order in
the expansion in external momenta. For a review of GXPT see \cite{GXPTrev}.}
of ``generalized chiral perturbation theory'' (GXPT) \cite{GXPTori}.
Meanwhile, they accept that the lower critical bound might be somewhat larger,
say $N_{\!f\,\mr{crit}}'\!=\!O(5)$, but the interesting question remains
whether there is any trace of the adjacent ``conformal window'' for lower
$\Nf$-values where QCD is in the confined phase and the chiral symmetry is
broken -- the latter most likely signaled by the lowest dimensional possible
order parameter already, i.e.\ though $\Sigma\!>\!0$ (distinctively).
If in this ``ordinary'' phase $\Sigma$ would indeed get substantially reduced
when $\Nf$ is increased by one unit, such a behaviour could be interpreted
as a hint that the real world with $\Nf\!=\!3$ is already ``close'' to the
lower bound $N_{\!f\,\mr{crit}}'$, as argued by the promoters of GXPT
\cite{GXPTrev}.
Hence, if the values quoted in (\ref{fpifinal}) would indeed represent the
ultimate truth for $\Nf\!=\!2$ (while still $F_\pi\!\simeq\!93\,\mr{MeV}$ or
$\Sigma\!\simeq\!(288\,\mr{MeV})^3$ for $\Nf\!=\!3$), this could be taken as an
indication that the real world is not too far away from the lower end of the
``conformal window''.
The important point is, of course, that such a hypothesis is accessible to
direct tests on the lattice%
\footnote{A simple test would be to measure $\ch(m)$ in QCD with $\Nf\!=\!4$
degenerate flavours and to show that the topological susceptibility curve has,
for $m\!\to\!0$, an asymptotic slope which is {\em less than half\/} of that
in the bottom part of Fig.\ 9. The ultimate test is, of course, to show that
--~despite the supposed correctness of (\ref{fpifinal})~-- one still finds
$F_\pi\!\simeq\!93\,\mr{MeV}, \Sigma\!\simeq\!(288\,\mr{MeV})^3$ in 2+1-flavour
QCD.},
and due to those the current body of evidence in favour of a relatively low
$N_{\!f\,\mr{crit}}'$ \cite{Mawhinney:1998jm,theofavour} will either be whiped
out or corroborated%
\footnote{Intuitively, one might think that the result (\ref{fpifinal}) of the
present analysis is strong evidence against this scenario, but things are not
that simple: If $N_{\!f\,\mr{crit}}'$ turns out so low that even $\Sigma_2$
is zero or marginal (i.e.\ not sufficiently large that the first term on the
r.h.s.\ of the GOR-relation dominates; cf.\ footnote 20), then the result
(\ref{fpifinal}) itself is affected, since in such a case $\Sigma$ may not
simply be read off from $\ch(m)$.}.
As an ironic twist one should note that the value (\ref{fpifinal}) which we
supposed as ``ultimate'' in this scenario and which pushed us into these
speculative thoughts is, by its numerical size, {\em anything but marginal\/}.

%%%%%%%%%%%%%%%%%%%%%%%%%%%%%%%%%%%%%%%%%%%%%%%%%%%%%%%%%%%%%%%%%%%%%%%%%%%%%%%

\section{Summary and Outlook}

In this paper I have confronted recent lattice data for the topological
susceptibility $\ch$ in $\Nf\!=\!2$ QCD as a function of the
(dynamical) quark mass with theoretical expectations.

The ultimate goal of these studies is to understand how QCD manages to reduce
the prominent $\th$-dependence of quenched QCD up to the point where it is
eliminated, if the chiral limit is taken.
Dynamical simulations, by themselves, provide little insight into the mechanism
behind this fascinating behaviour; what they can do, however, is to document
that the topological susceptibility gets suppressed (and hence the
$\th$-dependence reduced) if the dynamical quark masses are tuned sufficiently
small.
Pure theoretical thought, on the other hand, may provide interesting bounds
and it may yield information about some asymptotic behaviour, but typically
the information produced is qualitative, i.e.\ the relevant constants have to
be ``borrowed'' from elsewhere.

The approach taken here is to combine the lattice data with analytical insight.
This is complicated by the fact that the preferred ranges of quark masses are
quite different.
To be on safe analytical grounds one likes the quarks to be either sufficiently
light so that the flavour-singlet axial WT-identity enforces the linear
relationship (\ref{susclarge}) or sufficiently heavy so that the topological
susceptibility effectively coincides with its quenched counterpart.
The lattice data, on the other hand, happen to be precisely in the intermediate
regime which --~based on our preliminary knowledge on $\Sigma$ and $\ch_\infty$
in QCD~-- was estimated to be at quark masses of order $88\,\mr{MeV}$ in the
2-flavour theory and where the topological susceptibility follows neither one
of the two asymptotic patterns.

The key idea on which this article is based is that it is possible to combine
the leading physics resources of either regime without invoking specific model
assumptions.
This is done on the more general level of the finite-volume QCD partition
function $Z_\nu$ from which the topological susceptibility may be derived.
What enters the construction (\ref{Zcombi}) is the assumption that the two
mechanisms which suppress configurations with higher $|\nu|$ w.r.t\ those with
lower $|\nu|$ in the small- and the large-mass regime, respectively, operate
independently of each other.
For small enough quark masses the relevant mechanism is captured in the leading
order chiral Lagrangian (which in turn implements the constraint imposed by the
axial WT-identity), and the suppression gets manifest upon evaluating the
functional integral over the pseudo-Goldstone boson manifold.
For large enough quark masses instantons are the relevant degrees of freedom,
and an argument is pushed forward that the suppression of higher topological
sectors follows from a pure entropy consideration, if instantons are taken
incompressible.
To the extent that instanton interactions may be neglected and the relevant
physics is, on that side, indeed pure entropy, the hypothesis that the two
suppression mechanisms are independent is then correct by construction.

On the practical level, the outcome of this analysis is twofold.
First, for large enough box volumes the functional form (\ref{suscombi}) is
established as a well-motivated fitting curve, suitable for the entire range
of quark masses encountered in present and future studies of QCD with
$\Nf\!\geq\!2$ (and $\Nf\!<\!N_{\!f\,\mr{crit}}'$).
It contains only two fitting parameters, and both of them have a direct
interpretation in terms of (semi-)phenomenological quantities ($\Sigma$ and
$\ch_\infty$).
Second, the two different choices for the chiral finite-volume partition
function $Z_\nu^\mr{XPT}$ (i.e.\ (\ref{LSlarge}) vs.\ (\ref{LSall})) entering
the construction of $Z_\nu^\mr{QCD}$ allow one to quantitatively assess
--~within the validity of (\ref{Zcombi})~-- the impact of finite-volume effects
on the topological susceptibility.
The somewhat surprising outcome is that they are smaller than present days
statistical uncertainties for relatively small values of the LS-parameter $x$
already, say for $x\!\geq\!10$.

Making use of the functional form (\ref{suscombi}) the data by CP-PACS, UKQCD,
SESAM/TXL and the Pisa group have been analyzed.
The results for the asymptotic parameters $\Sigma$ and $\ch_\infty$ seem to
vary more than what one would have expected from the error bars of the
individual data points.
While the quenched topological susceptibility $\ch_\infty$ is in loose
agreement with both the estimate from the Witten-Veneziano relation and
previous direct determinations in the quenched theory, the result for the
low-energy constant $\Sigma\!=\!-\lim_{m\to0}\lim_{V\to\infty}\<\psb\ps\>$
outstrips the expected value by a factor 2.
Here the expectation relies on the GOR-relation and the lattice determination
of the non-strange quark mass; would one use a more traditional estimate of
$(m_u\!+\!m_d)/2$, the factor by which $\Sigma$ as determined from the
topological susceptibility curve exceeds our expectations would be larger.
Even restricting the set to contain nothing but ``high quality'' data (which
have passed certain cuts to ensure that effects due to finite lattice spacing
and finite box volume should be small) hardly changes the result of the
analysis -- the $\Nf$-fold asymptotic slope for $\Nf\!=\!2$, $\Sigma_2$, stays
high.

Two possible reasons for this excess have been discussed.
The simple explanation (supported by some inherent features of the data) is
that it is due to lattice artefacts.
The alternative one (which is mainly discussed because of it conceptual appeal
and its potential to initiate dedicated studies) is that $\Sigma_2$ is indeed
much larger than the GOR-relation suggests, indicating a larger-than-believed
$\Nf$-dependence of low-energy QCD.
In either case, it seems fair to say that at least in 2-flavour QCD chiral
symmetry is predominantly broken through a (distinctively nonzero) chiral
condensate \cite{Colangelo:2001sp}.

The implication is, of course, that further studies are needed to clarify the
situation (cf.\ \cite{furtherstudies}).
From the analysis one gathers that there are basically three reasonable
strategies for going more chiral: to simulate
({\em i\/}) at fixed $\be$,
({\em ii\/}) at fixed $a$,
({\em iii\/}) at fixed $x$,
whereas the possibility ({\em iv}) to simulate at fixed $\hat m$ seems to yield
more questionable results, since in this approach $\be$ gets so much relaxed
that the lattice gets intolerably coarse for the most chiral points.
While it is clear that in principle (i.e.\ if a continuum extrapolation
were possible) any of these possibilities would yield the same physical
answer, in practice the choice may make a sizable difference, and it is
important to keep in mind that the ``optimum'' choice may not only depend on
the regime of quark masses and box volumes one wishes to cover, but also on
the actions used.
In this respect it is clear that --~even if one cannot afford a continuum
extrapolation for all the data points constituting the $\ch\!=\!\ch(m)$
curve~-- a scaling study at least for a single quark mass value might help to
get a more quantitative assessment of discretization effects and also to
indicate which might be the most promising combination of actions in the gluon
and fermion sectors, respectively.
If progress turns out to be slow in the $\Nf\!=\!2$ system, studying QCD with
$\Nf\!=\!4$ might be an interesting digression -- not only because $\Sigma_4$
would tell us whether the ``conformal window'' is close, but mainly for the
practical reason that the ``transition regime'' occurs at higher quark mass
values and is therefore cheaper to study.
What we are ultimately interested in, however, are the values gained in a more
ambitious project, namely the simulation of 2+1-flavour QCD.

%%%%%%%%%%%%%%%%%%%%%%%%%%%%%%%%%%%%%%%%%%%%%%%%%%%%%%%%%%%%%%%%%%%%%%%%%%%%%%%

\bigskip\noindent{\bf Acknowledgements}\newline
It is a pleasure to acknowledge interesting and useful E-mail correspondence
with Ruedi Burkhalter.
In addition, I would like to thank the CP-PACS collaboration for allowing me
to show their results for the topological charge distribution on the smaller
lattice (Fig.~2).

%%%%%%%%%%%%%%%%%%%%%%%%%%%%%%%%%%%%%%%%%%%%%%%%%%%%%%%%%%%%%%%%%%%%%%%%%%%%%%%

\end{document}